\documentclass[11pt,english]{article}

\usepackage[T1]{fontenc}
\usepackage[margin=1in]{geometry}
\usepackage[latin9]{inputenc}
\usepackage{color}
\usepackage[dvipsnames]{xcolor}
\usepackage{enumerate}
\usepackage{enumitem} 
\usepackage{hyperref}
\usepackage{subcaption}
\definecolor{darkblue}{rgb}{0, 0, 0.5}
\hypersetup{
     colorlinks = true,
     citecolor = Maroon,
     urlcolor = darkblue,
     linkcolor = darkblue
}
\usepackage[normalem]{ulem} 

\usepackage{amsmath}
\usepackage{amsthm}
\usepackage{amssymb}
\usepackage{cases}
\usepackage{bm}
\usepackage[authoryear]{natbib}
\usepackage{setspace}
\usepackage{graphicx}
\usepackage{float}

\usepackage{macros}

\makeatletter
\theoremstyle{plain}
  \newtheorem{rem}{\protect\remarkname}

  \theoremstyle{plain}
  \newtheorem{assumption}{\protect\assumptionname}
  
\theoremstyle{plain}
\newtheorem{thm}{\protect\theoremname}
\newtheorem*{thm*}{\protect\theoremname}
\newtheorem{cor}{\protect\corname}
  \theoremstyle{plain}
  \newtheorem{lem}{\protect\lemmaname}[section]
    \newtheorem{lemMT}{\protect\lemmaname}
  \theoremstyle{remark}

\makeatother

\usepackage{babel}
  \providecommand{\assumptionname}{Assumption}
  \providecommand{\claimname}{Claim}
  \providecommand{\lemmaname}{Lemma}
  \providecommand{\remarkname}{Remark}
\providecommand{\theoremname}{Theorem}
  \providecommand{\examplename}{Example}
\providecommand{\corname}{Corollary}

\title{Bias correction for quantile regression estimators\footnote{This paper supersedes ``Conditional quantile estimators:
A small sample theory''.}}
\author{
\setcounter{footnote}{1}
        Grigory Franguridi\thanks{
		Center for Economic and Social Research (CESR), University of Southern California.
		Email: \href{franguri@usc.edu}{franguri@usc.edu}
	}
	\and Bulat Gafarov\thanks{
	Department of Agricultural and Resource Economics, University of California, Davis. Research for this paper was supported by  the USDA National Institute of Food and Agriculture, Hatch project S-1072: U.S. Agricultural
Trade and Policy in  an Uncertain Global Market Environment. 
Email: \href{bgafarov@ucdavis.edu}{bgafarov@ucdavis.edu}
	}
	\and Kaspar W\"uthrich\thanks{
	Department of Economics, University of Michigan; CESifo. Email: \href{kasparwu@umich.edu}{kasparwu@umich.edu}
	}
}

\begin{document}

\maketitle

\begin{abstract}

    \linespread{1.2}

We study the bias of classical quantile regression and instrumental variable quantile regression estimators.
While being asymptotically first-order unbiased, these estimators can have non-negligible second-order biases.
We derive a higher-order stochastic expansion of these estimators using empirical process theory. 
Based on this expansion, we derive an explicit formula for the second-order bias and propose a feasible bias correction procedure that uses finite-difference estimators of the bias components.
The proposed bias correction method performs well in simulations. 
We provide an empirical illustration using Engel's classical data on household food expenditure.

\medskip


\medskip


\end{abstract}
\newpage

 \onehalfspacing

\section{Introduction}\label{sec:intro}

Many interesting empirical applications of classical quantile regression (QR) \citep{koenker1978regression} and instrumental variable quantile regression (IVQR) \citep{chernozhukov2005iv,chernozhukov2006instrumental} feature small sample sizes, which can arise either as a result of a limited number of observations or when estimating tail quantiles, or both \citep[e.g.,][]{chernozhukov2005annals,elsner2008increasing,chernozhukov2011inference,adrian2016covar,adrian2019vulnerable}.
QR and IVQR estimators are nonlinear and can thus exhibit substantial biases in small samples.
In this paper, we theoretically characterize these biases and develop a feasible bias correction procedure.

To study the biases, we start by deriving a higher-order stochastic expansion of the classical QR and exact IVQR estimators.\footnote{We define exact IVQR estimators as estimators that exactly minimize a norm of the sample moment conditions. Such estimators can be obtained using mixed-integer programming (MIP) methods \citep[e.g.,][]{chen2018exact,zhu2019learning}. See Appendix \ref{app:LP_MILP}.}
Such an expansion is needed because the higher-order terms contribute nonzero biases while the first-order term does not.
We derive explicit expressions and uniform (in the quantile level) rates for the components in the expansion, building on the empirical process arguments of \citet{ota2019quantile}.
This expansion can be thought of as a refined Bahadur-Kiefer (BK) representation of the estimator that decomposes the nonlinear component into terms up to the order $O_p\bigPar{n^{-1}}$ and a  $O_p\bigPar{n^{-5/4}\sqrt{\log n}}$ remainder. We also derive a uniform BK representation for generic IVQR estimators after a feasible 1-step Newton correction (see  Appendix \ref{app: Stochastic expansion of general quantile regression estimators}).

Using the stochastic expansion, we study the bias of QR and exact IVQR estimators.
We derive a bias formula based on the leading terms up to order $O_p\bigPar{n^{-1}}$ in the expansion, which we refer to as the \emph{second-order (asymptotic) bias}. This approach of focusing on the moments of the leading terms in the stochastic expansions is standard in the literature.\footnote{See, for example, \citet{nagar1959bias,newey2004higher,kato2012asymptotics,galvao2016smoothed, kaplan2017smoothed,hahn2023efficient}, among others.} 
The second-order bias formula provides an approximation of the actual bias that yields a feasible correction.
Our results explicitly account for the bias due to nonzero sample moments at the estimator. 
Our proof strategy is different from the generalized function heuristic used in the existing literature \citep{phillips1991shortcut,lee2017second,lee2018second}, which does not account for all the terms in the second-order bias (see Section \ref{sec: Bias formula for exact estimators} and Appendix \ref{app: Illustration of approximate bias formula in univariate case} for further discussion and examples).  The missing terms can be important bias contributors in practice, as we document in the empirical application in Section \ref{sec: empirical application}. 

A feasible bias correction procedure then follows from the second-order bias formula.
We propose finite-difference estimators of all the components in the formula.
These estimators admit higher-order expansions that allow us to select bandwidth rates. In particular, our finite-difference estimator for the Jacobian coincides with \citet{powell1986censored}'s classical estimator, and the bandwidth rate we derive coincides with the AMSE optimal bandwidth choice in \citet{kato2012asymptotic}. 
We show that the resulting (analytically) bias-corrected estimator has zero second-order bias.
This result is in contrast to the commonly-used bootstrap bias correction approaches \citep{horowitz2001bootstrap}, which may not capture the higher-order bias terms of quantile estimators  \citep[see][]{knight2003second}.

We evaluate the performance of our bias correction procedure in a Monte Carlo simulation study. 
The simulations show that the theoretical (infeasible) bias formula describes well the second-order bias of classical QR and exact IVQR.
We find that the proposed feasible bias correction can effectively reduce the bias in many cases.
The gains from bias correction are particularly prominent in settings with large bias such as for the IVQR estimators under endogeneity. 
The impact of the bias correction on the root MSE (RMSE) is rather small and ambiguous.

We illustrate the bias correction approach by revisiting the relationship between food expenditure and income based on the original \citet{engel1857} data \citep[e.g.,][]{koenker1982robust,koenker2001quantile}.\footnote{The QR approach was used more recently to analyze changes in heterogeneity in the income elasticity over time of various consumer spending categories in \cite{taylor2009consumer}.} 
Our results highlight the importance of bias correction in empirical applications with small sample sizes. 
Specifically, we find that the second-order bias of classical QR is empirically relevant: it can be larger than 50\% of the standard error, which is substantial given that this is the second-order bias.

\paragraph{Roadmap.}
 The remainder of the paper is organized as follows. Section \ref{sec:setup_background} describes the model and the estimators.
 Section \ref{sec: Asymptotic theory for bias correction} provides our main theoretical results. 
 Section \ref{sec:MC} presents the Monte Carlo simulation results.
 Section \ref{sec: empirical application} contains the empirical application.
 Section \ref{sec:conclusion} concludes. All the proofs and some additional details are given in the Appendix.

\section{Model and estimators}
\label{sec:setup_background}
 
Consider a setting with a continuous outcome variable $Y$, a $(k\times 1)$ vector of covariates $W$, and a $(k\times 1)$ vector of instruments $Z$. 
We assume throughout that $k$ is fixed.
Every observation $(Y_i,W_i,Z_i)$, $i=1,\dots,n$, is jointly drawn from a distribution $P$. 
We assume that $(Y_i,W_i,Z_i)$ is i.i.d., and we will sometimes suppress the index $i$ to lighten up the notation. 
The parameter of interest  $\thetaTrue\in\Theta\subset \R^k$ is defined as a solution to the following unconditional quantile moment restrictions,
\begin{equation}
\expect[(1\{Y\leq W^{\prime}\thetaTrue\}-\tau) Z ] = 0, \quad \tau\in (0,1).\label{eq:unconditionalMoment}
\end{equation}
We consider two cases: (i) classical QR, where $ Z=W$ \citep{koenker1978regression}, and (ii) linear IVQR, where $Z\neq W$ in general \citep{chernozhukov2006instrumental,chernozhukov2008instrumental}. 

The classical QR estimator of $\thetaTrue$ is a solution to the following convex minimization problem,
\begin{equation}
     \hat\theta_{\tau,QR} \in \operatorname{argmin}_{\theta\in\Theta}  \emp \rho_\tau(Y-W^\prime \theta ), \label{eq:QR_check_function_minimization}
\end{equation}
where $\rho_\tau(u)=u(\tau-1\{u<0\})$ is the check function \citep{koenker2005quantile} and $\emp$ denotes the sample average, i.e., the expectation with respect to the empirical measure. For IVQR, we consider estimators that exactly minimize the $p$-norm of the sample moments,
 \begin{equation}
\thetaLp \in \operatorname{argmin}_{\theta\in \Theta} ||\gthat(\theta)||_p,  \label{eq:GMM_Lp}   
\end{equation}
where $p\in[1,\infty]$ and $\gthat(\theta)\bydef  \emp  (1\{Y\leq W^{\prime}\theta\}-\tau)Z$. This class of exact IVQR estimators includes GMM, which corresponds to $p=2$ as in \citet{chen2018exact} for just-identified models, and  the estimator proposed by \citet{zhu2019learning}, which corresponds to $p=\infty$. 
The cases $p=1$ and $p=\infty$ have computationally convenient mixed integer linear programming (MILP) representations, while the MILP formulation for $p=2$ has many more decision variables. In our Monte Carlo simulations, we use $p=1$ for computational convenience (see Appendix \ref{app:LP_MILP}).

We use the notation $\gt(\theta) \bydef \expect(1\{Y\leq W^{\prime}\theta\}-\tau) Z$ for the unconditional moment restrictions as a function of $\theta\in\Theta$, and write $G(\theta)\bydef \partial_\theta \expect Z1\{Y\leq W^{\prime}\theta\}=\partial_\theta g_{\tau}(\theta)$ for its derivative. 
We maintain the following standard identification assumptions.
\begin{assumption}[Identification] \label{ass:identification}
\text{ }
\begin{enumerate}
    \item $\thetaTrue$ is the unique solution to $\gt(\theta)=0$ over a compact set $\Theta \subset \R^k$, and $\thetaTrue$ is in the interior of $\Theta$ for all $\tau\in [\varepsilon,1-\varepsilon]$ for some $\varepsilon>0$.\label{ass:identification_global}
    \item The Jacobian $G(\thetaTrue)$ has full rank for all $\tau\in (0,1)$.\label{ass:identification_full_rank}
\end{enumerate}
\end{assumption}

As noted by \citet{chernozhukov2006instrumental}, ``compactness [of the parameter space $\Theta$] is not restrictive in micro-econometric applications'' (p.502). Throughout the paper, we use the short notation $G$ for $G(\thetaTrue)$ whenever it does not lead to ambiguity.

We impose the following smoothness assumptions on the conditional density and its derivatives. Such assumptions are standard in the literature on higher-order properties of quantile estimators \citep[e.g.,][]{ota2019quantile}.
\begin{assumption}[Conditional density]\label{ass:density}
The conditional density of $Y_i$ given $(W_i,Z_i)$, $f_{Y}(y|w,z)$, exists, is a.s.\ three times continuously differentiable on $supp(Y)$, and there exists a constant $\bar{f}$ such that $|f^{(r)}_{Y}(y|w,z)|\le \bar{f}$ for all $(y,w,z)\in supp(Y)\times supp(W)\times supp(Z)$, where $r=0,1$ and $f_Y^{(r)}(\cdot|w,z)$ is the $r$-th derivative of $f_Y(\cdot|w,z)$.
\end{assumption}
In our theoretical analysis of the bias, we will often work with a related object, the conditional density $f_{\varepsilon_{\tau}}(e|W,Z)  \bydef f_{Y}(e+W'\thetaTrue|W,Z)$ of the quantile residual $ \varepsilon_{\tau}  \bydef Y- W^\prime\thetaTrue$.

\label{REPLY:R1.1.part1}
Finally, we assume that the regressors and the instruments have bounded higher-order moments.
\begin{assumption}[Regressors and instruments]\label{ass:regressors_instruments} There exists   constants $m<\infty$ and $\gamma\geq6$  such that   $\expect |W_j|^\gamma \leq m$  and $\expect |Z_j|^\gamma \leq m$   for all $j=1,\dots,k$.
\end{assumption}
Assumption \ref{ass:regressors_instruments} guarantees the existence of all relevant moments of the terms involving $W$ and $Z$ in the higher-order derivatives of the moment conditions and the bias correction (e.g., $\expect Z_\ell W_j W_q  W_r$). 
The power parameter $\gamma$ (as we show below) determines the upper bound on the rate at which the norm of the sample moment functions converges to zero --- higher $\gamma$ implies faster convergence to zero.

\section{Asymptotic theory for bias correction}
\label{sec: Asymptotic theory for bias correction}
To derive a bias correction procedure, we follow the approach of \cite{nagar1959bias} and focus on the bias of the leading terms in the asymptotic stochastic expansion of the estimator.

\subsection{Stochastic expansion of quantile regression estimators}
\label{sec:BK_IVQR}

The classical first-order asymptotic theory for quantile regression estimators \citep[e.g.,][]{koenker1978regression,angrist2006quantile,chernozhukov2006instrumental,kaplan2017smoothed,kaido2021decentralization} is based on the following leading term, 
\begin{equation}
   \hat\xi_\tau \bydef   \thetaTrue  - G^{-1}(\thetaTrue)  \emp  Z\bigPar{1\{Y\le W^\prime\thetaTrue\} - \tau    }. \label{eq:definitionOfTheta_1}
\end{equation}
For correctly specified models, $\hat\xi_\tau$ is an infeasible unbiased estimator of $\thetaTrue$.
However, because feasible quantile estimators are nonlinear, they generally have a nonzero higher-order bias.
The following theorem provides a characterization of the terms in a stochastic expansion of $\thetahat$ up to order $O_p\bigPar{n^{-1}}$ (ignoring logarithmic terms).

To state the result, we introduce some additional notation. 
For $\theta\in\Theta$, define the auxiliary functions $g^\circ(\theta) \bydef \expect1\{Y\leq W^{\prime}\theta\}Z$, 
$B^\circ_n(\theta)    \bydef  \sqrt{n}  (\emp 1\{Y\leq W^{\prime} \theta \} Z  - g^\circ (\theta))$, and $B_n(\theta)  \bydef B^\circ_n(\theta) - B^\circ_n(\thetaTrue)$.\footnote{The processes $B_n(\theta) $ and $B^\circ_n(\theta) $ take values in the space $\ell^{\infty}(\Theta)$ of bounded functions on $\Theta$.} 
Also, denote the Hessian of the $j$-th moment function $g_j$ by
  $\partial_\theta G_j(\theta)  \bydef \partial_{\theta } \partial_{ \theta}  g_j(\theta) $.
 Finally, for any $x\in \R^{k}$, denote by $x^\prime  \partial_\theta G(\theta) x$ the vector with components $x^\prime  \partial_\theta G_j(\theta) x$, $j=1,\dots,k$.

\begin{thm} \label{thm:stochasticExpansion}  
Suppose that Assumptions~\ref{ass:identification}--\ref{ass:regressors_instruments} hold. 
Consider $\thetahat = \thetaLp$ obtained from program \eqref{eq:GMM_Lp} for some $p\in [1,\infty]$ or  $\thetahat = \hat\theta_{\tau,QR}$.
Then 
\begin{equation*}
\thetahat  = \hat\xi_\tau+ 
     G^{-1} (\thetaTrue)\bigBra{\gthat(\thetahat) - \frac{B_n(\thetahat)}{\sqrt{n}}  - \frac{1 }{2} (   \hat\xi_\tau-\thetaTrue)^\prime \partial_\theta G(\thetaTrue)  (\hat\xi_\tau-\thetaTrue) } + R_{n,\tau},
\end{equation*}
where
\begin{align}
    \sup_{\tau \in [\varepsilon,1-\varepsilon]}\|\gthat(\thetahat)\| &=
    \begin{cases}
         O_p\bigPar{\frac{ 1  }{ n }}, \text{ if } \thetahat=\thetaQR,\\
         O_p\bigPar{\frac{ \log n  }{ n^{1-\frac{2}{\gamma}} }}, \text{ if } \thetahat=\thetaLp, 
    \end{cases} \label{eq:ghat_supbound} \\
    \sup_{\tau \in [\varepsilon,1-\varepsilon]}\| B_n(\thetahat) \|&= O_p\bigPar{\frac{\sqrt{\log n}}{n^{1/4}}} \label{eq:Bn_supbound}, \\
        \sup_{\tau \in [\varepsilon,1-\varepsilon]}\| \hat\xi_\tau-\thetaTrue\|
        &=   O_p\bigPar{\frac{ 1}{n^{1/2}}},\label{eq:DonskerLinTerm}\\
    \sup_{\tau \in [\varepsilon,1-\varepsilon]}\|R_{n,\tau}\|&=O_p\bigPar{\frac{\sqrt{\log n}}{n^{5/4}}}. \notag 
\end{align}
\end{thm}

The proof of Theorem \ref{thm:stochasticExpansion} builds on the empirical process arguments in Lemma 3 of \citet{ota2019quantile} and uses the maximal inequality in Corollary 5.1 of \citet{chernozhukov2014gaussian}.

The rates in Theorem \ref{thm:stochasticExpansion} are uniform in the quantile level $\tau$. 
Uniformity is important in theory and practice because QR and IVQR methods are particularly powerful when used to analyze the entire quantile process.

Classical QR and linear IVQR are motivated by the linearity of the true conditional quantile function. In practice, linearity may be restrictive, especially when it is imposed at each $\tau\in (0,1)$. While the result in Theorem \ref{thm:stochasticExpansion} remains valid if linearity fails, the interpretation is more complicated in this case. Specifically, if the true conditional quantiles are nonlinear, then the expansion in Theorem \ref{thm:stochasticExpansion} applies to the pseudo-true value defined via moment condition \eqref{eq:unconditionalMoment}. For classical QR, this pseudo-true value can be interpreted as the minimizer of a weighted mean-squared error loss function \citep{angrist2006quantile}. Note further that the result in Theorem \ref{thm:stochasticExpansion} holds for every fixed $\tau\in (0,1)$. Thus, if linearity holds at a given $\tau$, we can interpret the expansion at this $\tau$ under the correct specification without requiring linearity at the other quantile levels.

The expansion in Theorem \ref{thm:stochasticExpansion} can be thought of as a refined Bahadur-Kiefer (BK) expansion. 
Different from standard BK expansions, we do not bundle together all the higher-order terms \citep[as opposed to, for example,][]{zhou1996direct,ota2019quantile}.
Notice that the dominant nonlinear term in the BK expansion, $ n^{-1/2}G^{-1}B_n(\thetahat)$,  has order  $O_p\bigPar{ n^{-3/4}\sqrt{\log n} }$ (Equation \eqref{eq:Bn_supbound}). 
Theorem 3 in \citet{knight2002comparing} shows that for classical QR with discrete covariates, this term converges in distribution to a zero mean random process.
Therefore, we explicitly extract the higher-order terms up to order $O_p\bigPar{n^{-1}}$ (ignoring logarithmic terms) from the BK remainder.  
 As we will show in the following sections, these higher-order terms admit feasible counterparts.
\begin{rem}[Alternative approach for deriving stochastic expansions]
\citet{portnoy2012nearly} proposed an alternative approach for deriving a stochastic expansion of classical QR estimators.  
This approach yields bounds on the precision of a nonlinear Gaussian approximation of order $O_p\bigPar{n^{-1}\log^{5/2}n}$. 
As we will show, the expansion in Theorem \ref{thm:stochasticExpansion} yields a bias formula for both QR and IVQR estimators that admits a feasible implementation.
The results in \citet{portnoy2012nearly} are specific to classical QR, and it is not clear to us whether these results can be used for bias correction using a Nagar-style approach. \qed
\end{rem}

\begin{rem}[General IVQR estimators]
While we focus on exact IVQR estimators in the main text, the results in Theorem \ref{thm:stochasticExpansion} can be used to obtain a uniform BK expansion for general 1-step corrected IVQR estimators. See Appendix \ref{app: Stochastic expansion of general quantile regression estimators} for details. \qed
\end{rem}
 
\subsection{Bias formula for exact estimators}
\label{sec: Bias formula for exact estimators}

Following common practice \citep[e.g.,][]{nagar1959bias,kaplan2017smoothed}, for a generic estimator $\hat\gamma$, we define the \emph{second-order bias}  $\bias(\hat\gamma)$ as the bias of the leading terms in the stochastic expansion of $\hat\gamma$ up to the order $O_p\bigPar{n^{-1}}$.
This second-order bias can be interpreted as an approximation of the actual bias that works with arbitrarily high probability in large samples.

Before stating the result, we observe that under our Assumption \ref{ass:density}, the moment condition \eqref{eq:unconditionalMoment} is equivalent to 
\begin{equation*}
    \expect[(1\{- Y\leq W^{\prime}(-\thetaTrue)\}-(1-\tau)) Z ] = 0.
\end{equation*}
Thus, we can characterize the QR and IVQR estimators using the moment function
$g_\tau^*(\theta) \bydef \expect[(1\{- Y\leq W^{\prime}\theta\}-(1-\tau)) Z ]  $
with sample analog $\gthat^\ast(\theta)$. 
The following theorem characterizes the second-order bias in terms of $\gthat(\hat\theta) $ and $ \gthat^*(-\hat\theta) $.
As in Section \ref{sec:BK_IVQR}, we define $\partial_\theta G_j(\theta)  \bydef \partial_{\theta } \partial_{ \theta}  g_j(\theta)  $ for all $\theta\in\Theta$.

\begin{thm}\label{thm:bias} Suppose that Assumptions~\ref{ass:identification}--\ref{ass:regressors_instruments} hold. Consider $\thetahat = \thetaLp$ obtained from program \eqref{eq:GMM_Lp} for some $p\in [1,\infty]$ or  $\thetahat = \hat\theta_{\tau,QR}$.
Then the second-order bias is
\begin{equation}
 \bias(\thetahat) =   G^{-1}(\thetaTrue) \bigBra{ \frac{1}{2}\expect  \bigPar{  \gthat(\thetahat) - \gthat^*(-\thetahat)  }  -   \frac{\kappa_{\tau}}{n}   - \frac{1}{2n }  Q^\prime vec( \Omega_\tau ) },\label{eq:biasFormula}
\end{equation}
where
\begin{align*}
    \kappa_{\tau }  &\bydef \bigPar{\tau -\frac{1}{2}}  \expect f_{ \varepsilon_\tau} (0 |W, Z) Z W^{\prime} G^{-1}Z, \\
    \Omega_\tau &\bydef  \operatorname{Var} [Z(1\{Y\le W' \thetaTrue\}-\tau)],\\
    Q &\text{ is a matrix with columns } Q_j \bydef \operatorname{vec}\bigBra{ ( G^{-1})^\prime   \partial_\theta G_j(\thetaTrue)    G^{-1} }, \,\, j=1,\dots,k.
\end{align*}
\end{thm}

The second-order bias formula \eqref{eq:biasFormula} has three components.

The first component, $ G^{-1}\expect  \bigPar{  \hat g_\tau(\thetahat) - \hat g_\tau^*(-\thetahat)  }/2 $, captures the bias from the sample moments not being zero at the estimator. 
This term is not equal to zero in general, as we illustrate based on a simple example in Appendix \ref{app: Illustration of approximate bias formula in univariate case}.

The second component, $n^{-1}G^{-1} \kappa_\tau $, appears because of the discontinuity in the sample moment functions.
The term $\kappa_\tau$ reflects the dependence between the sample moments and the linear influence of a single observation on $\thetahat$. 
Notice that the first two components combined correspond to the second-order bias of the terms $G^{-1} (\thetaTrue)(\gthat(\thetahat) -  n^{-1/2} B_n(\thetahat)) $  in Theorem \ref{thm:stochasticExpansion}.

The last component, $(2n)^{-1}G^{-1}Q^\prime vec(\Omega)$, stems from the non-uniformity of the conditional distribution of $Y$ given $(W,Z)$.
This term corresponds to the term $ G^{-1} (\thetaTrue) (\hat\xi_\tau-\thetaTrue)^\prime \partial_\theta G(\thetaTrue)  (\hat\xi_\tau-\thetaTrue)/2 $  in Theorem \ref{thm:stochasticExpansion}.
Similar terms are typically present in most nonlinear estimators with nonzero Hessian of the score function \citep[see, for example,][]{rilstone1996secondorder}.

\begin{figure}[ht]
\begin{center}
\caption{Comparison to exact formula in univariate case}
\includegraphics[width=0.7\textwidth]{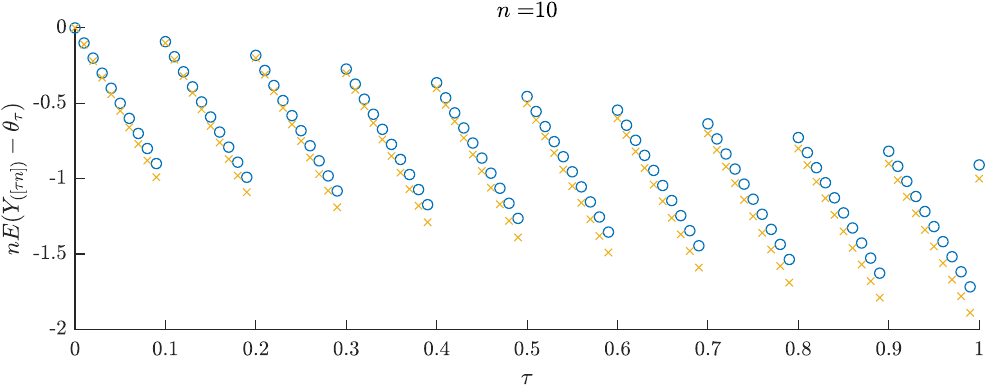}
\label{fig:exact_asymptotic_uniform_quantile}
\end{center}
 \footnotesize{\textit{Notes:} Exact (circles) and second-order (crosses)  biases, scaled by $n$, as functions of quantile level $\tau$ for $\thetahat = Y_{(\intPart{\tau n})}$, where $Y\sim \text{Uniform}(0,1)$, $n=10$.}

\end{figure}

To illustrate the approximate bias formula, consider an order statistic of $Y\sim \text{Uniform}(0,1)$ (corresponding to our framework with $Z=W=1$), for which an exact bias formula is available \citep[e.g.,][]{ahsanullah2013introduction}. 
We show in Appendix \ref{app: Illustration of approximate bias formula in univariate case} that the precision of the second-order bias formula in this case is  $O\bigPar{n^{-2}}$, which is smaller than the order of the remainder term in the stochastic expansion of Theorem \ref{thm:stochasticExpansion}.
Figure \ref{fig:exact_asymptotic_uniform_quantile} illustrates the precision of the asymptotic formula by comparing it to the actual bias of the order statistic. 

It is interesting to compare our results to the higher-order bias analysis of non-smooth estimators based on the generalized functions heuristic \citep[e.g.,][]{phillips1991shortcut}.
In recent work, \citet{lee2017second,lee2018second} derived a second-order bias formula for classical QR and IVQR under the assumption that the sample moments are zero at the estimator so that the first term of the bias formula vanishes. 
We show in Appendix \ref{app: Illustration of approximate bias formula in univariate case} that this term is non-negligible even in simple cases (see also Figure~\ref{fig:empirical_application} in Section~\ref{sec: empirical application}).

\subsection{Feasible bias correction} \label{sec:feasibleCorrection}

The bias formula suggests the following feasible bias-corrected estimator,
\begin{equation}
  \hat\theta_{bc}= \thetahat -  \frac{1}{2}\hat{G}^{-1} \bigBra{\gthat(\thetahat)-\gthat^*(-\thetahat)}  + \frac{1}{n} \hat G^{-1} \bigBra{   \hat \kappa_{\tau }  + \frac{1}{2} \hat Q^\prime vec( \hat \Omega ) }   , \notag 
\end{equation}
where $\hat G$, $\hat\kappa_{\tau}$, $\hat Q$, and $\hat\Omega$ are estimators of $G$, $\kappa_{\tau}$, $Q$, and $\Omega$, respectively, satisfying the following consistency requirement.
\begin{assumption}[Consistency of component estimators]
\label{ass:consistent_estimators}
The estimators $\hat G$, $\hat\kappa_{\tau}$, $\hat Q$, and $\hat\Omega$ are consistent for $G$, $\kappa_{\tau}$, $Q$, and $\Omega$, respectively. Moreover, $\hat G-G = o_p\bigPar{(n^{1/3}\log n)^{-1}}$.
\end{assumption}
In Section \ref{sec: finite difference estimators of bias components}, we propose finite-difference estimators for which Assumption \ref{ass:consistent_estimators} holds under Assumptions \ref{ass:identification}--\ref{ass:regressors_instruments}.
Assumption \ref{ass:consistent_estimators} could also be verified for other nonparametric estimators of the bias components. 

The next theorem shows that the second-order bias of the bias-corrected estimator is zero.
\begin{thm}\label{thm:biascorrection} Suppose that Assumptions \ref{ass:identification}--\ref{ass:consistent_estimators} hold. Consider $\thetahat = \thetaLp$ obtained from program \eqref{eq:GMM_Lp} for some $p\in [1,\infty]$ or  $\thetahat = \hat\theta_{\tau,QR}$. 
 Then the feasible bias correction eliminates the second-order bias, $\bias \bigPar{ \hat\theta_{bc} } = 0.$
\end{thm}

Note that the requirement that $\hat G-G = o_p\bigPar{(n^{1/3}\log n)^{-1}}$ in Assumption \ref{ass:consistent_estimators} is necessary to ensure that the contribution of the product of the sample moments, which are $O_p\bigPar{n^{-1+2/\gamma}\log n}$ with $\gamma\geq 6$ (as required by Assumption \ref{ass:regressors_instruments}) by Theorem \ref{thm:stochasticExpansion}, and the estimation error in $G^{-1}$ can be omitted in computing the second-order bias.
This condition is only required for IVQR estimators. For classical QR estimators, the sample moments are of order $O_p\bigPar{n^{-1}}$, so that consistency of $\hat G$ at any rate of convergence suffices for Theorem \ref{thm:biascorrection}.

\subsection{Finite difference estimators of bias components}
\label{sec: finite difference estimators of bias components}
To implement the bias correction, we need estimators of  $G$, $\kappa_{\tau}$, $  Q$, and $ \Omega$ that satisfy Assumption \ref{ass:consistent_estimators}.
The variance matrix $\Omega$ can be estimated using the analogy principle,
\begin{equation*}
     \hat \Omega_{\tau} \bydef \emp [Z(1\{Y\le W'\thetahat\}-\tau)-\emp Z(1\{Y\le W'\thetahat\}-\tau)]^2.
\end{equation*}
All other bias components take the form of derivatives.
Therefore, we leverage our theoretical results on the properties of the sample moments to develop a unified finite-difference framework for estimating these components.

Under Assumptions~\ref{ass:identification} and \ref{ass:density}, the Jacobian is $G=\expect f_{\varepsilon_{\tau}}(0|W,Z)ZW'$ and the Hessian consists of gradients of the components of $G$, i.e. $\partial_\theta G_{i,j}(\thetaTrue)=\expect f^{(1)}_{\varepsilon_{\tau}}(0|W,Z) Z_i W_j W $, where $i,j=1,\dots,k.$
This suggests the following analog estimators.

 The $(i,j)$-th component of $G$ can be estimated using \citet{powell1986censored}'s estimator
\begin{equation}
    \hat G_{i,j}=   \emp \left[ \frac{1\{Y\le W'\thetahat+ h_{1,n}\} - 1\{Y\le W'\thetahat- h_{1,n}\} }{2h_{1,n}}Z_i W_j \right], \label{eq:G-hat-ij}
\end{equation}
where $h_{1,n} \to 0$ is a bandwidth. Denote by $e_\ell$ the $\ell$-th unit vector in $\R^k$, where $\ell=1,\dots, k$. The derivative of the $(i,j)$-th component of $G$ in the direction $e_\ell$ (i.e. the second partial derivative of $\gt$) can be estimated as the symmetric first difference of \eqref{eq:G-hat-ij}, 
\begin{align*}
        \widehat{({\partial_\theta G}_{i,j})}_\ell &= \emp \left[ \frac{1\{Y\le W'\thetahat+ h_{2,n}\} - 2 \cdot 1\{Y\le W'\thetahat\} +  1\{Y\le W'\thetahat- h_{2,n}\}}{  h_{2,n}^2} Z_i W_j W_\ell \right],
\end{align*}
where $h_{2,n} \to 0$ is a (potentially different) bandwidth. For $\kappa_{\tau}$, the finite difference sample analog is 
\begin{equation*}
    \hat \kappa_{\tau} = \bigPar{\tau-\frac{1}{2}} \emp \left[ \frac{1\{Y\le W'\thetahat+ h_{3,n}\} - 1\{Y\le W'\thetahat- h_{3,n}\} }{2h_{3,n}} Z W^{\prime} \hat G^{-1}Z \right],
\end{equation*}
where   $h_{3,n} \to 0$ is a bandwidth. Finally, $Q$ can be estimated by the sample analog matrix $\hat Q$ with columns
\begin{equation*}
     \hat Q_j \bydef vec\bigBra{ ( \hat G^{-1})^\prime \widehat{\partial_\theta G_j} \hat G^{-1} }, \quad j=1,\dots,k,
\end{equation*}
where $ \widehat{\partial_\theta G_j} $ is the matrix with elements $\widehat{({\partial_\theta G}_{i,j})}_\ell$ for $i,\ell=1,\dots,k$.

The next lemma establishes the consistency of the estimators of the bias components. It implies that these estimators satisfy the high-level conditions in Assumption \ref{ass:consistent_estimators}.
Moreover, it provides \emph{nearly remainder-optimal} bandwidth rates, i.e., the rates that yield the fastest convergence rates of the stochastic remainder terms of the corresponding stochastic expansions (up to logarithmic terms). 

\begin{lemMT}\label{thm: finite difference}
Suppose that Assumptions~\ref{ass:identification}--\ref{ass:regressors_instruments} hold. 
Then the nearly remainder-optimal bandwidth rates are $h_{1,n} \propto n^{-1/5}$, $h_{2,n} \propto n^{-1/7}$, $h_{3,n} \propto n^{-1/5}$, and, under these bandwidth rates,
\begin{align*}
     \hat G &= G + O_p\bigPar{\frac{\sqrt{\log n}}{n^{2/5}}},\\
     \widehat{({\partial_\theta G}_{i,j})}_\ell &= {({\partial_\theta G}_{i,j})}_\ell + O_p\bigPar{\frac{\sqrt{\log n}}{  n^{2/7} }},\\
      \hat Q_j &=  Q_j + O_p \bigPar{  \frac{\sqrt{\log n}}{n^{2/7}} }, \\
    \hat \kappa_{\tau} &=  \kappa_{\tau}     + O_p\bigPar{\frac{\sqrt{\log n}}{n^{2/5}}},\\
    \hat \Omega_{\tau} &= \Omega + O_p \bigPar{ \frac{1}{\sqrt{n}} }.
\end{align*}
Moreover, the convergence rate for $\hat G$ is uniform in $\tau\in[\varepsilon,1-\varepsilon]$.
\end{lemMT}
 
Proposition 1 in \cite{kato2012asymptotic} shows that the remainder rate for $\hat{G}$ in Lemma \ref{thm: finite difference}, $h_{1,n} \propto n^{-1/5}$, is the AMSE-optimal rate.\footnote{
We conjecture that analogous AMSE-optimality results could be established for estimators $\widehat{({\partial_\theta G}_{i,j})}_\ell$ and $\hat Q_j $, but leave this extension for future work.}

To implement the finite difference estimators in practice, one needs to choose constants in addition to the bandwidth rates. A natural approach would be to use non-parametric methods such as cross-validation to estimate the optimal bandwidth parameter. However, such non-parametric methods typically require large sample sizes \citep[e.g.][]{simonoff1996smoothing}, which makes them impractical for our purposes. Another common approach, which we take here, is to propose rule-of-thumb choices that account for the scale of the data. To be robust to heavy-tailed distributions, we suggest using a robust measure of dispersion. Specifically, we propose a rule-of-thumb choice in form $h_{1,n}=A_{G}\cdot n^{-1/5}$, $h_{2,n}=A_Q\cdot n^{-1/7}$, $h_{3,n}=A_\kappa\cdot n^{-1/5}$, where, for all $j\in \{G,Q,\kappa\}$,
\begin{equation*}
A_j= \tilde{A}_j \cdot 1.48 \cdot \widehat{\operatorname{MAD}}_\tau ,
\end{equation*}
 and $\widehat{\operatorname{MAD}}_\tau$ is the estimated median absolute deviation of the $\tau$-quantile residuals. The constant $1.48$ is chosen because for the normal distribution 1.48 times the median absolute deviation is equal to the standard deviation.\footnote{\cite{chernozhukov2013inference} suggest using $\operatorname{IQR}/1.35$ as a robust estimate of the scale parameter. For the normal distribution, this choice coincides with $1.48 \operatorname{MAD}$.} We suggest choosing $\tilde{A}_G=\tilde{A}_\kappa=2 $ and $\tilde{A}_Q=1.5 $ based on our simulation evidence, as we describe in more detail in Section \ref{subsec:BandwithChoice}.

\section{Simulation evidence}
\label{sec:MC}

In this section, we evaluate the performance of our feasible bias correction procedure in a Monte Carlo simulation study. 

\subsection{Simulation design}

We consider data-generating processes (DGPs) inspired by the simulations in \citet{andrews2016conditional}. 
The outcome is generated according to the following location-scale model
\begin{align}
    Y_i=   W_i+( 0.5 + W_i)U_i,\quad i=1,\dots,n, \notag 
\end{align}
where $W_i =  \Phi(\tilde{W}_i)$, $Z_i = \Phi(\tilde{Z}_i)$, $U_i = F_U^{-1}\left(\Phi(\tilde{U}_i)\right)$, $(\tilde{W}_i,\tilde{Z}_i, \tilde{U}_i)\sim N(0,\Sigma)$, $\Sigma_{11}=\Sigma_{22}=\Sigma_{33}=1$, $\Sigma_{23}=0$, and $\Phi$ is the standard normal CDF.
Hence, in all the designs, both the regressors and the instruments are $\operatorname{Uniform} [0,1]$. We consider six DGPs that differ with respect to the error distribution $F_U$ and whether or not $W$ is exogenous.

\begin{center}
\begin{tabular}{ l  l  l}
\hline
\hline
 DGP1 (Uniform, exogenous) & $F_U(u)=\int_{-\infty}^u 1\{t\in[0,1]\}dt$ & $\Sigma_{12}=1$, $\Sigma_{13}=0$  \\ 
 \hline
 DGP2 (Triangular, exogenous) & $F_U(u)=\int_{-\infty}^u 2 t 1\{t\in[0,1]\}dt$& $\Sigma_{12}=1$, $\Sigma_{13}=0$   \\  
 \hline
 DGP3 (Cauchy, exogenous) & $F_U(u)=\int_{-\infty}^u \frac{1}{\pi(1+(4 t)^2)}dt$ & $\Sigma_{12}=1$, $\Sigma_{13}=0$\\ 
 \hline
 DGP4 (Uniform, endogenous) &$F_U(u)=\int_{-\infty}^u 1\{t\in[0,1]\}dt$ & $\Sigma_{12}=0.75$, $\Sigma_{13}=0.25$\\
\hline
 DGP5 (Triangular, endogenous) &$F_U(u)=\int_{-\infty}^u 2 t 1\{t\in[0,1]\}dt$ & $\Sigma_{12}=0.75$, $\Sigma_{13}=0.25$\\
\hline
 DGP6 (Cauchy, endogenous) &$F_U(u)=\int_{-\infty}^u \frac{1}{\pi(1+(4 t)^2)}dt$  & $\Sigma_{12}=0.75$, $\Sigma_{13}=0.25$\\

 \hline
 \hline
\end{tabular}
\end{center}

In Appendix \ref{sec:additionalFigures}, we consider two additional DGPs to assess the impact of the strength of the instrument.

\subsection{Bandwidth choice}\label{subsec:BandwithChoice}

An important practical issue is the choice of the bandwidths $h_{1,n}$, $h_{2,n}$, and $h_{3,n}$. We use the (data-dependent) rule-of-thumb bandwidth choice described in Section \ref{sec: finite difference estimators of bias components}. To implement this rule-of-thumb bandwidth choice, we need to choose the constants $\tilde{A}_j$, $j\in \{G,Q,\kappa\}$. We found that across the DGPs considered in our simulations, $\tilde{A}_G=\tilde{A}_\kappa=2$ and $\tilde{A}_Q=1.5$ perform well in terms of bias across all designs, especially when the bias is large.\footnote{If more information about the DGP is available, it is possible to further refine the tuning parameter. For example, under DGP1 where $Q=0$, choosing smaller values for $A_Q$ results in a better alignment with the infeasible formula. However, absent such information, we consider the proposed rule of thumb to be a reasonable compromise when one cares about the performance across different designs, especially those with larger biases (e.g., DGP3--DGP6).}

We compare the performance of the feasible bias correction based on the rule-of-thumb bandwidth choice to the performance of the corresponding infeasible bias correction based on the true $G$, $Q$, and $\kappa$. For IVQR, the exact formulas are not available. Therefore, we use numerically computed values based on 10 million observations and tuning choice  ${A}_G={A}_Q={A}_\kappa=1$.

\subsection{Results}

\begin{figure}[H]
	\caption{Bias (multiplied by $n$) before and after correction for DGP1--DGP3}

	\begin{center}
	\begin{subfigure}[b]{0.3\textwidth}
		\includegraphics[width=\textwidth]{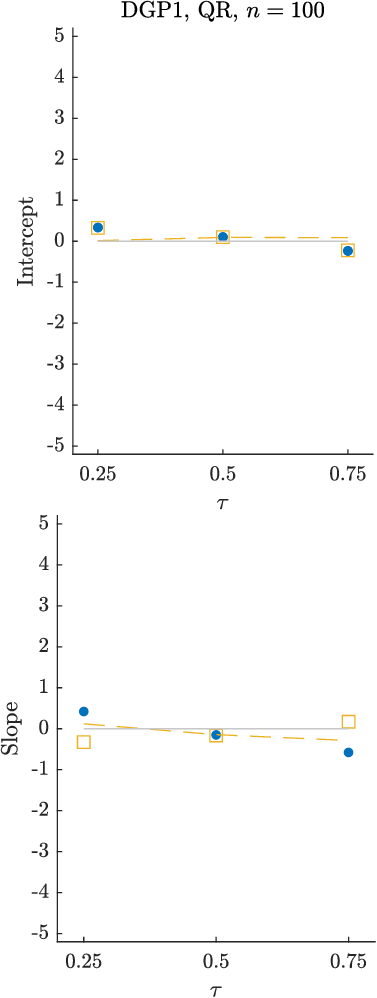}
		
	\end{subfigure}
	\begin{subfigure}[b]{0.3\textwidth}
		\includegraphics[width=\textwidth]{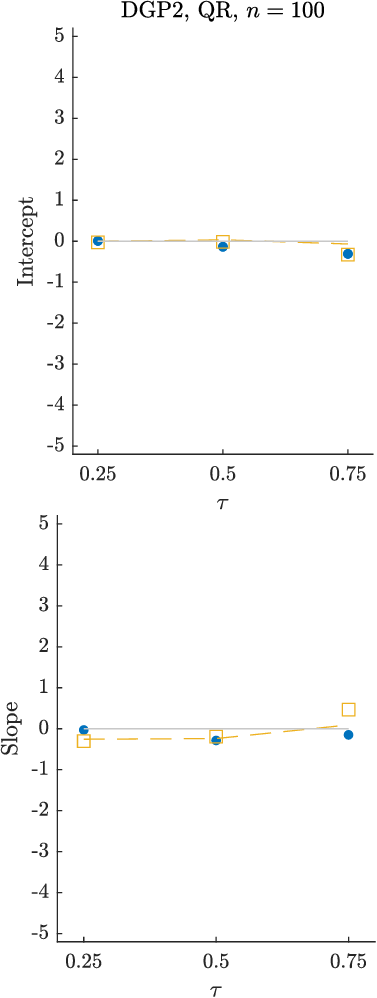}
	
	\end{subfigure}
	\begin{subfigure}[b]{0.3\textwidth}
		\includegraphics[width=\textwidth]{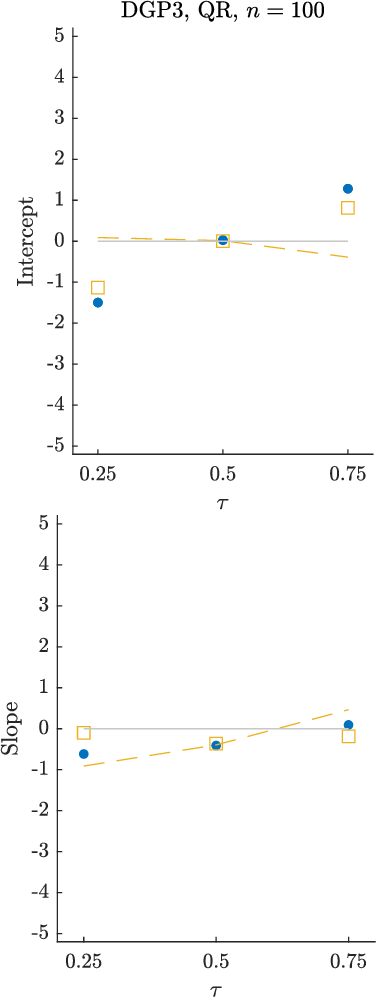}
	
	\end{subfigure}
		\end{center}
	 \footnotesize{\textit{Notes:} The panels display the bias (multiplied by $n$) of the intercept and the slope for classical QR without bias correction (blue dots), QR with feasible bias correction based on the rule-of-thumb bandwidth (gold squares), and QR with infeasible bias correction (gold dashed line) for DGP1--DGP3. All results are based on 5,000 simulation repetitions.}
 \label{fig:bias_dgp13}
\end{figure}

\begin{figure}[H]
	\caption{Bias (multiplied by $n$) before and after correction for DGP4--DGP6}

	\begin{center}
	\begin{subfigure}[b]{0.3\textwidth}
		\includegraphics[width=\textwidth]{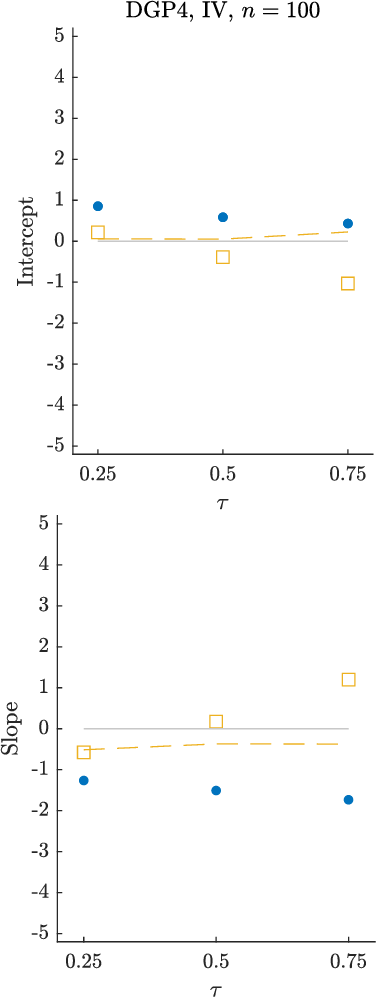}
		
	\end{subfigure}
	\begin{subfigure}[b]{0.3\textwidth}
		\includegraphics[width=\textwidth]{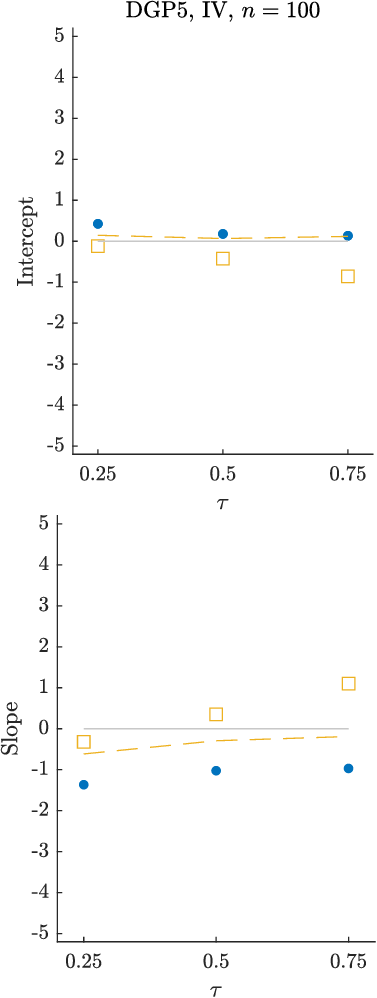}
	
	\end{subfigure}
	\begin{subfigure}[b]{0.3\textwidth}
		\includegraphics[width=\textwidth]{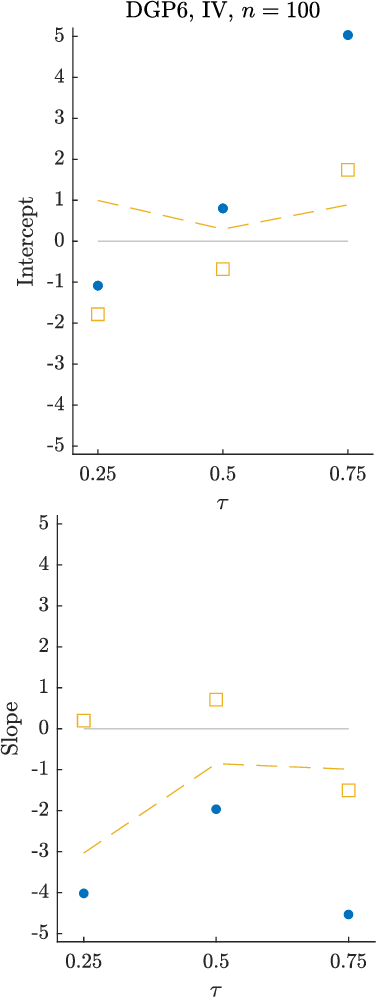}
	
	\end{subfigure}
		\end{center}
	 \footnotesize{\textit{Notes:} The panels display the bias (multiplied by $n$) of the intercept and the slope for IVQR (implemented via the MILP formulation in Appendix \ref{app:LP_MILP}) without bias correction (blue dots), IVQR with feasible bias correction based on the rule-of-thumb bandwidth (gold squares), and IVQR with infeasible bias correction (gold dashed line) for DGP4--DGP6. All results are based on 5,000 simulation repetitions. The infeasible bias correction is based on the feasible formula applied to a simulated sample of 10,000,000 observations.}
 \label{fig:bias_dgp13_iv}
\end{figure}

We focus on the performance of bias correction for $\tau\in \{0.25, 0.5, 0.75\}$. Figure \ref{fig:bias_dgp13} shows the impact of bias correction for the exogenous DGP1--DGP3 with $n=100$. Figure \ref{fig:bias_dgp13_iv} shows the corresponding results for the endogenous DGPs (DGP4--DGP6). We use classical QR of $Y$ on $W$, implemented via the linear programming formulation in Appendix \ref{app:LP_MILP}, for DGP1--DGP3 and IVQR, implemented using the MILP formulation in Appendix \ref{app:LP_MILP}, for DGP4--DGP6.
 
The main findings can be summarized as follows. First, the QR estimators based on the exogenous designs (DGP1--DGP3) exhibit lower biases than the IVQR estimators based on the corresponding endogenous designs (DGP4--DGP6). For both estimators, the biases tend to be the largest for the designs with heavy tails (DGP3 and DGP6). Second, the feasible bias correction based on the rule-of-thumb bandwidth choice reduces the bias of QR and IVQR estimators in many cases. The bias reductions are the most notable when the biases of the original estimators are large, which is when bias correction is most needed.
Finally, the infeasible bias correction reduces the bias in most cases, underscoring the usefulness of the proposed theory.
See Appendix \ref{sec:additionalFigures} for additional simulation evidence on the impact of the sample size and instrument strength on the performance of bias correction.

\begin{figure}[H]
	\caption{RMSE comparison of raw and bias-corrected estimators }\label{fig:RMSEuniform}

	\begin{center}
	\begin{subfigure}[b]{0.45\textwidth}
		\includegraphics[width=\textwidth]{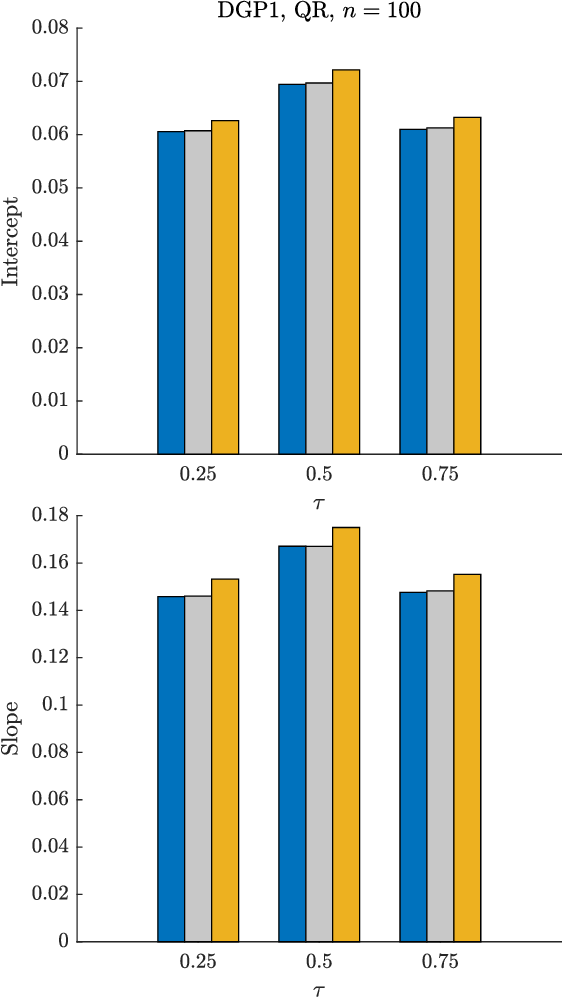}
		\caption{}
	\end{subfigure}
	\begin{subfigure}[b]{0.45\textwidth}
		\includegraphics[width=\textwidth]{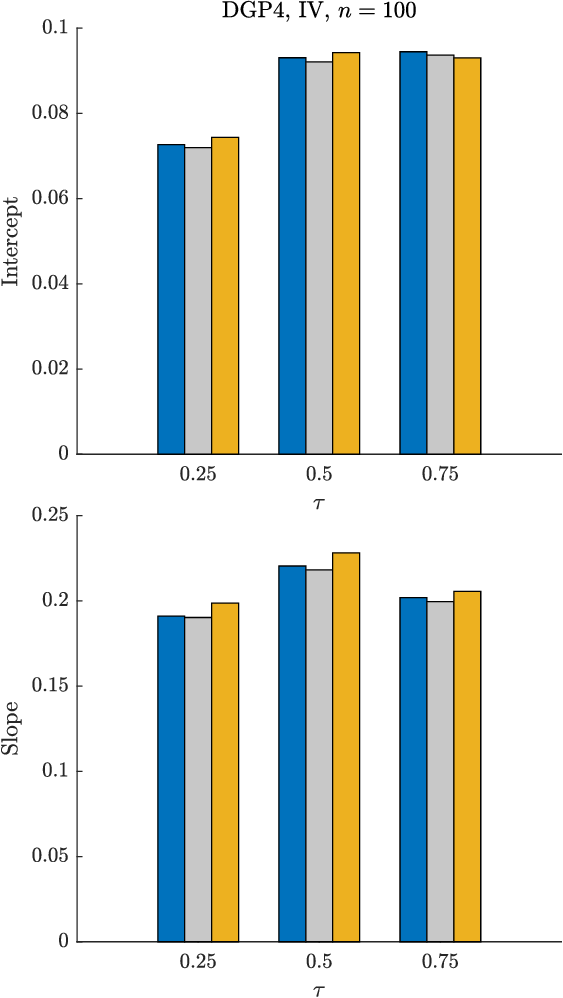}
	\caption{}
	\end{subfigure}

		\end{center}
	 \footnotesize{\textit{Notes:}
   The panels compare the RMSE for estimators without bias correction (blue), with infeasible bias correction (grey) and with feasible bias correction based on the rule-of-thumb bandwidth choice (gold) for (a) DGP1, classical QR  and (b) DGP4, IVQR. All results are based on 5,000 simulation repetitions. 	 
	 }
	\label{fig:rmse14}
\end{figure}

 Next, we investigate the impact of bias correction on the RMSE of the estimators. The proposed bias correction approach is designed to reduce the bias but is not theoretically guaranteed to reduce the RMSE. Figure \ref{fig:rmse14} reports the results for DGP1 and DGP4 with $n=100$. 
 Overall, the impact of bias correction on the RMSE is rather small. While the infeasible bias correction can slightly increase the RMSE for DGP1, it decreases the RMSE across all quantile levels for DGP4. The feasible bias correction with the rule-of-thumb bandwidth slightly increases the RMSE in most cases. 
Appendix Figures \ref{fig:rmse25} and \ref{fig:RMSEstable} show the results for the other DGPs, and Appendix Figure \ref{fig:mad} shows the corresponding results for the Mean Absolute Deviation, an alternative measure of risk. 

\begin{figure}[H]
	\caption{Confidence interval coverage before and after correction for DGP1 and DGP4}
	\begin{center}
		\begin{subfigure}[b]{0.3\textwidth}
		\includegraphics[width=\textwidth]{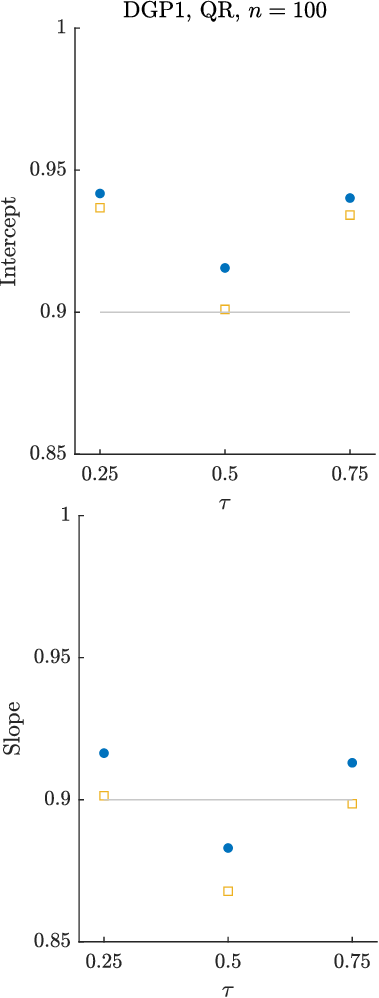}
	\caption{}
	\end{subfigure}
	\begin{subfigure}[b]{0.3\textwidth}
		\includegraphics[width=\textwidth]{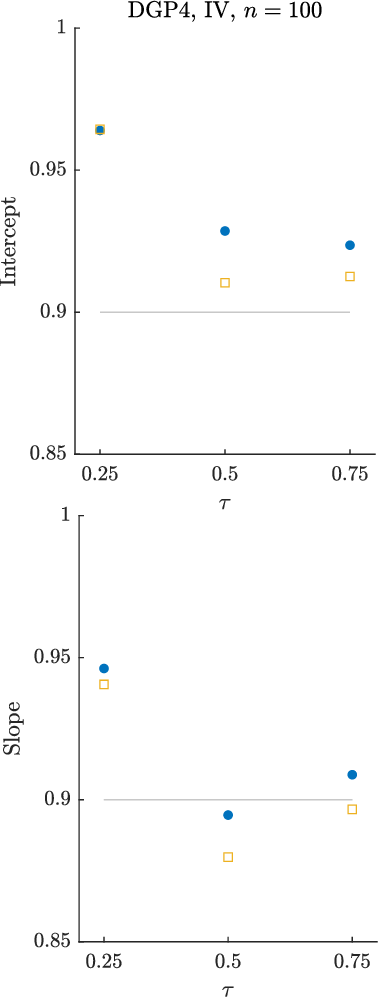}
	\caption{}
	\end{subfigure}

		\end{center}
	 \footnotesize{\textit{Notes:} 
  The panels display the coverage probability of the $90\% $ confidence intervals for the intercept and the slope  without bias correction (blue dots) and  with the  feasible bias correction based on the rule-of-thumb bandwidth choice (gold squares) for DGP1 (classical QR) and DGP4  (IVQR). All results are based on 5,000 simulation repetitions.}
	\label{fig:coverageDGP1DGP4}
\end{figure}

Finally, we study the impact of the bias correction on the coverage probability of the standard confidence intervals.
To focus on the impact of bias correction, we use the same standard errors based on the rule-of-thumb bandwidth for the original and the bias-corrected estimator. As a result, by construction, the bias correction does not affect the length of the confidence intervals.
Figure \ref{fig:coverageDGP1DGP4} shows the empirical coverage for DGP1 and DGP4 before and after bias correction. 
The feasible bias correction based on the rule-of-thumb bandwidth leads to higher coverage accuracy in the majority of cases but can lead to some undercoverage at the median. Thus, while bias correction may not improve the RMSE in small samples, it can lead to more accurate inferences. The results for the other DGPs are in Appendix Figures \ref{fig:coverageDG25} and \ref{fig:coverageDG36}, and Appendix Figure \ref{fig:coverageDGP1_DGP4_n200} presents the results for $n=200$.

\section{Empirical application}
\label{sec: empirical application}
The second-order bias matters most in applications with small sample sizes. 
We therefore illustrate our bias correction approach using the classical dataset of \citet{engel1857}, analyzed by \citet{koenker1982robust} and \citet{koenker2001quantile}, among others.
The data contain information on annual income and food expenditure (in Belgian francs) for $n=235$ Belgian working-class households and are obtained from the \texttt{R} package \texttt{quantreg} \citep{quantreg_package}. 
One feature of these data is the growing dispersion of the outcome variable (food expenditure) as a function of the regressor (income) \citep{koenker2001quantile}, which is similar to our Monte Carlo designs.
We divide the values of income and food expenditure by $1000$ so that the unit of measurement becomes a thousand Belgian francs. 
This makes the scale of intercept and slope parameters comparable. 

\begin{figure}[H]
	\caption{Quantile regression of annual food expenditure on income}

	\begin{center}
	\begin{subfigure}[b]{0.45\textwidth}
		\includegraphics[width=\textwidth]{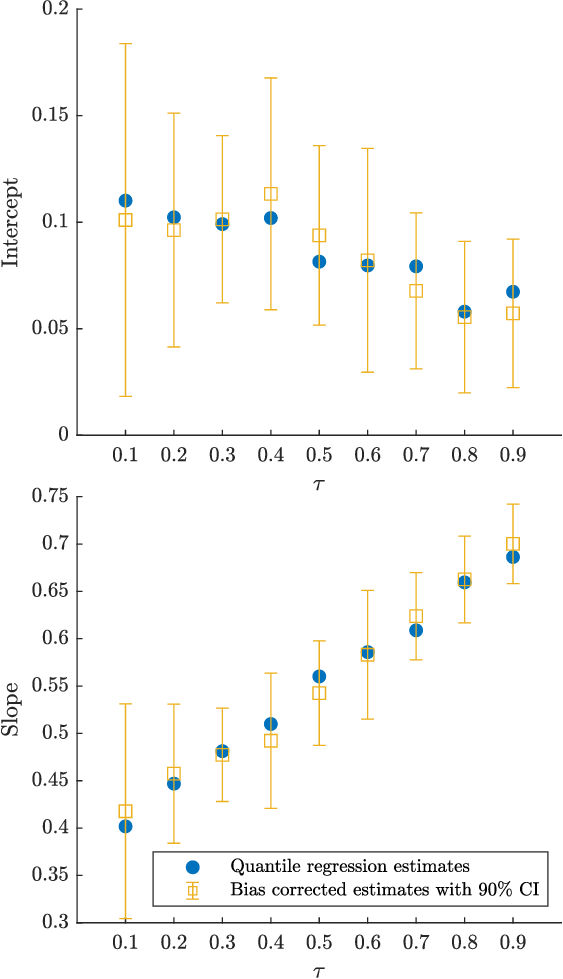}
		\caption{Impact of bias correction}
	\end{subfigure}
	\begin{subfigure}[b]{0.45\textwidth}
		\includegraphics[width=\textwidth]{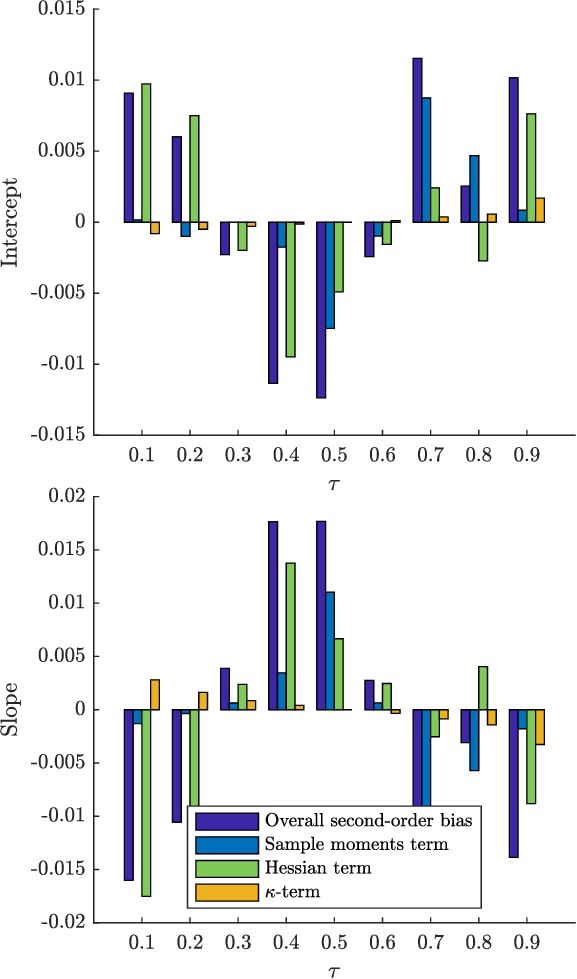}
	\caption{Composition of second-order bias}
	\end{subfigure}

		\end{center}
	 \footnotesize{\textit{Notes:} Panel (a) compares the classical QR estimates (blue dots) to the bias-corrected QR estimates (gold squares) with 90\% confidence intervals (bars). The bias correction is based on the rule-of-thumb bandwidth choice with $(\tilde{A}_G,\tilde{A}_Q,\tilde{A}_\kappa)=(2,1.5,2)$. Panel (b) shows the contributions of the different bias components to the overall second-order bias.
	 }
	\label{fig:empirical_application}
\end{figure}

We estimate classical QRs of food expenditure ($Y$) on income ($W$) and a constant (blue dots). The bias-corrected estimates (gold squares) are obtained using the recommended rule-of-thumb bandwidth choice with $(\tilde{A}_G,\tilde{A}_Q,\tilde{A}_\kappa)=(2,1.5,2)$. Figure \ref{fig:empirical_application} presents the results. 
Panel (a) compares the classical and the bias-corrected QR estimates. 
The results suggest that the impact of bias correction is more pronounced around the median and in the tails.
The magnitude of the differences between the original and bias-corrected estimates can be larger than 50\% of the standard errors, which is substantial given that we are focusing on the second-order bias.

Panel (b) shows the individual contributions of the different components to the overall second-order bias. 
We can decompose the bias correction term as follows:
\begin{align*}
  \thetahat-\hat\theta_{bc}=   \underbrace{\frac{1}{2}\hat{G}^{-1} \bigBra{\gthat(\thetahat) -\gthat^*(-\thetahat)}}_{\text{(i)}}
  -\underbrace{  \frac{1}{n} \hat G^{-1}    \hat \kappa_{\tau }}_{\text{(ii)}}
  -\underbrace{    \frac{1}{2n}G^{-1}\hat Q^\prime vec( \hat \Omega )}_{\text{(iii)}} .
\end{align*}
The main takeaway from the bias decomposition is that while all three components play a role, the sample moment term (i) and especially the Hessian term (iii) can be large and account for most of the bias. The $\kappa$-term (ii) is smaller overall and only matters in the tails.

\section{Conclusion}
\label{sec:conclusion}
 
We demonstrate that classical QR and IVQR estimators can exhibit a non-negligible second-order bias.
We characterize this bias theoretically and use this characterization to derive a novel analytical bias correction method.
The proposed feasible bias correction reduces the bias of QR and IVQR estimators across a variety of settings at a very low computational cost.
However, there is scope for further improving its performance when the sample size is very small and the instruments are weak. 
The simulation performance of the infeasible bias correction based on the population version of our theoretical bias formula suggests that exploring alternative feasible bias correction approaches is a promising direction for future research. For example, one could consider regularization approaches or explore imposing parametric assumptions to improve the estimation of the bias components.

\section*{Acknowledgments}
We are grateful to the Editor (Xiaohong Chen), the Associate Editor, and three anonymous referees, as well as Victor Chernozhukov, Zheng Fang, Dalia Ghanem, Jiaying Gu, Marc Henry, Keisuke Hirano, Nail Kashaev, Roger Koenker, Vladimir Koltchinskii, Michal Kolesar, Simon Lee, Blaise Melly, Hyungsik Roger Moon, Hashem Pesaran, Joris Pinkse, Wolfgang Polonik, Stephen Portnoy, Geert Ridder, Andres Santos, Davide Viviano, Yuanyuan Wan, and seminar participants at UC Berkeley, UC Davis, UC Los Angeles, University of Toronto, and USC for valuable comments. All errors and omissions are our own.

\bibliographystyle{ecta}
\bibliography{main}
 
\newpage  
\part*{Online appendix}
\appendix
 
\setcounter{page}{1}

\section{Bahadur-Kiefer representation, proofs} \label{sec:ap_zEstimator}

\subsection{Auxiliary results for generic IVQR estimators}
\label{app:Auxiliary results for generic IVQR estimators}

\begin{lem}\label{lem:differentiability_g(theta)}
Under Assumptions \ref{ass:density} and \ref{ass:regressors_instruments}, $g_\tau(\theta)$ is three times continuously differentiable in $\theta$.  
\end{lem}
\begin{proof}
By definition, $g_\tau(\theta)=\expect (1\{Y\leq W^{\prime}\theta\}-\tau) Z=\expect ( \expect (F_{Y}(W^{\prime}\theta|W,Z)-\tau) Z)$.
The result then follows from the dominated convergence theorem.
\end{proof}

\begin{lem} \label{lem:genericStochasticExpansion}
Suppose Assumptions \ref{ass:density} and \ref{ass:regressors_instruments} hold.
Then for any estimator $\thetahat $  such that
$ \sup_{\tau \in [\varepsilon,1-\varepsilon]}\| \hat\theta_\tau -  \thetaTrue  \|= O_p\bigPar{{r^{-1}_n}}$ for some sequence $r_n \to \infty$, we have a representation    
\begin{align}
    \hat{ g}_\tau(\thetahat ) &=   \frac{1}{\sqrt{n}} B^\circ_n(\thetaTrue) + \tau(\expect Z-\emp{Z}) + \frac{1}{\sqrt{n}}  B_n(\thetahat) \notag \\
    &+   G(\thetaTrue) (\thetahat-\thetaTrue)+   \frac{1 }{2} (   \thetahat-\thetaTrue)^\prime \partial_\theta G(\thetaTrue) (\thetahat-\thetaTrue) + O_p\bigPar{ \frac{1}{r^3_n} }, \label{eq:sampleMomentsExpansion} 
\end{align}
where the remainder rate is uniform in $\tau\in[\varepsilon,1-\varepsilon]$.

\end{lem}

\begin{proof} 
By definition,
\begin{align*}
   \gthat(\thetahat )& = \emp 1\{Y\leq  
      W^{\prime} \thetahat    \} Z - \tau \emp Z  \\
      & =\frac{1}{\sqrt{n}} B^\circ_n(\thetahat) +  g^\circ(\thetahat) - \tau \emp Z\\
      & = \frac{1}{\sqrt{n}}  B^\circ_n(\thetaTrue) + \frac{1}{\sqrt{n}} B_n(\thetahat)   + \tau (\expect Z-\emp{Z}) +  \gt(\thetahat).
\end{align*}
By Lemma~\ref{lem:differentiability_g(theta)}, $\gt(\cdot)$ is three times continuously differentiable. 
Since $\theta$ is restricted to a compact set $\Theta$, the norm of the third derivative is bounded on $\Theta$.
The Taylor theorem implies that there exist a neighborhood of $\thetaTrue$ such that for any $\theta$ in the neighborhood,
\begin{align*}
 \gt(\theta )  =  G(\thetaTrue)(\theta-\thetaTrue) +  \frac{1 }{2} (   \theta-\thetaTrue)^\prime \partial_\theta G(\thetaTrue) (\theta-\thetaTrue) + R(\theta),
 \end{align*}
 where $R(\theta) = O\bigPar{  \|\theta-\thetaTrue\|^3 }$  uniformly in $\tau$. 
 Then \eqref{eq:sampleMomentsExpansion} follows immediately because $\thetahat$ is a uniformly consistent estimator.
 \end{proof}

Now let us study the large sample behavior of the term $B_n(\thetahat)$ in \eqref{eq:sampleMomentsExpansion}.

\begin{lem}\label{lem:stochasticHolder} Suppose that Assumptions \ref{ass:density} and \ref{ass:regressors_instruments}  hold. For any pair of estimators $\thetahat$ and $\thetahat^\ast$ such that $ \sup_{\tau \in [\varepsilon,1-\varepsilon]}\| \thetahat^\ast- \thetahat\|= O_p\bigPar{{r^{-1}_n}}$ for some sequence $r_n \to \infty$, we have
\begin{equation*}
    B_n(\thetahat)- B_n(\thetahat^\ast)=
    O_p\bigPar{ \sqrt\frac{{\log r_n}}{r_n} } +o_p \bigPar{  \frac{\log r_n}{ n^{1/3}} } \quad \text{uniformly in } \tau\in[\varepsilon,1-\varepsilon].
\end{equation*}

\end{lem}

\begin{proof} The proof relies on the arguments in \citet[][Lemma 3]{ota2019quantile} adapted to our setting.
The idea is to verify the conditions of Lemma 1 of \citet{ota2019quantile}, which follows from Corollary 5.1 in \citet{chernozhukov2014gaussian} and use this corollary to prove the desired result.

\medskip

Since $ \sup_{\tau \in [\varepsilon,1-\varepsilon]}\| \thetahat^\ast- \thetahat\|= O_p\bigPar{{r^{-1}_n}}$, we have $P\bigPar{ \sup_{\tau \in [\varepsilon,1-\varepsilon]}\| \thetahat^\ast- \thetahat\|\le {M_n}/{r_n}}\to 1$ for any sequence $M_n\to\infty$.  
Consider the functions 
\begin{align*}
    &f_{\theta,h,\alpha}:\, (y,w,z) \mapsto \left(1\{ y- w^\prime \theta \le w^\prime h \} -1\{ y- w^\prime \theta \le 0\}\right) \alpha^\prime z
\end{align*}
that constitute the function class 
\begin{align*}
    \mathcal{F}_n &= \bigg\{ f_{\theta,h,\alpha}:  \,\, \theta\in\Theta, \|h\|\leq \frac{M_n}{r_n}, \|\alpha\|= 1 \bigg\},
\end{align*}
where $M_n$ is a sequence such that $M_n\to \infty$ and $M_n/r_n\to 0$. Let $\eG_n$ be the standard empirical process operator on $  \mathcal{F}_n$ with the data $(Y_i,W_i,Z_i)$, $i=1,\dots,n$. 
By Assumption \ref{ass:regressors_instruments},  $\mathcal{F}_n$ admits an envelope $F(y,w,z)\equiv \|z\|$. 

\medskip

Let us now verify the conditions in Lemma 1 of \citet{ota2019quantile}.
First, since
\begin{align*}
    \left(1\left\{ Y- W^\prime \theta \le W^\prime h \right\}
     -1\left\{ Y- W^\prime \theta \le 0\right\}\right)^2 = 1\left\{\min(0,W'h) < Y- W^\prime \theta \le \max(0,W'h) \right\},
\end{align*}
we obtain
\begin{align*}
    \expect f_{\theta,h,\alpha}^2(Y,W,Z) &= \E\left[ \left(1\big\{ Y- W^\prime \theta \le W^\prime h \big\}
     -1\big\{ Y- W^\prime \theta \le 0\big\} \right)\alpha'Z \right]^2 \\
      &\le  \E \|Z\|^2 \cdot \E 1\bigg\{\min(0,W'h) < Y- W^\prime \theta \le \max(0,W'h) \bigg\} \\
     &\le  k m^{2/\gamma} \cdot \E 1\bigg\{\min(0,W'h) < Y- W^\prime \theta \le \max(0,W'h) \bigg\} \\
     &=k m^{2/\gamma} \cdot \E \bigg( \E\left[ \left| F_{Y}(W'h+W^\prime \theta|W,Z) - F_{Y}( W^\prime \theta|W,Z) \right| \,\bigg|\,W,Z\right]\bigg)\\
     &\le k m^{2/\gamma} \cdot \expect\bigg( \E\left[  |W'h| \sup_y f_{Y}(y|W,Z) \, \bigg| \,W,Z\right] \bigg) \leq k^2 m^{4/\gamma} \bar{f}\|h\| =  O\bigPar{\frac{M_n}{r_n}},
\end{align*}
where we used Assumptions \ref{ass:density} and the inequality
\begin{equation*}
    \E \|Z\|^2 \le k \max_{j=1,\dots,k} \E |Z_j|^2 \le k  \max_{j=1,\dots,k} \left( \E |Z_j|^\gamma \right)^{2/\gamma}
\end{equation*}
in conjunction with Assumption \ref{ass:regressors_instruments}.

Therefore, the  variance parameter  of the process is
\begin{equation*}
    \sigma_n^2 \bydef \sup_{f\in\mathcal{F}_n} \expect f^2(Y,W,Z) = O\bigPar{\frac{M_n}{r_n}}.
\end{equation*}

Second, using Lemma 2 in \cite{ota2019quantile}, we have $$\expect\max_{1\leq i\leq n } F^2(Y_i,W_i,Z_i)=\expect \max_{1\leq i\leq n } \|Z_i\|^2 = o(n^{2/\gamma}). $$

Third, because the function class $\mathcal{F}_n$ is a VC class \citep{vapnik1971uniform} with the   envelope $\|Z\|$, there exist constants $A$ and $V$ independent of $n$ such that the standard entropy bound
\begin{align*}
    \sup_Q N\left( \mathcal{F}_n,\|\cdot\|_{Q,2},  \eta \|Z\|_{Q,2} \right) \le (A / \eta)^V, \text{ for all } \eta\in(0,1]
\end{align*}
holds \citep[e.g.,][Section 2.6]{van1996weak}. Here the supremum is taken over all finitely discrete measures $Q$ and $\|\cdot\|_{Q,2}$ is the $L^2(Q)$ norm.

Finally, applying Lemma 1 in \citet{ota2019quantile}, we obtain
\begin{eqnarray}
    &&\expect \sup_{\theta\in\Theta, \|h\|\leq \frac{M_n}{r_n}, \|\alpha\|= 1}\|\eG_n f_{\theta,h,\alpha}\|  \lesssim  \sqrt{V \sigma_n^2 \log (A m/ \sigma_n ) } +  \frac{V \sqrt{\expect \max_{1\leq i\leq n } \|Z_i\|^2 }}{\sqrt{n}} \log (A m/\sigma_n) \notag \\
    &&=  O\bigPar{  \sqrt\frac{\log r_n}{r_n} } +o \bigPar{  \frac{\log r_n}{n^{\frac{\gamma-2}{2\gamma}}} },\label{eq:useful_maximal_inequality}
\end{eqnarray}
where the last equality holds by choosing $M_n\to \infty$ sufficiently slowly.
Note that the right-hand side of this equation does not depend on $\tau$.
Consequently, by the definition of the norm and equation \eqref{eq:useful_maximal_inequality}, 
\begin{align*}
     \sup_{\tau\in[\varepsilon,1-\varepsilon]} \|B_n(\thetahat)- B_n(\thetahat^*)\|  &=   \sup_{\tau\in[\varepsilon,1-\varepsilon]} \max_{\|\alpha\|=1} \bigg| \eG_n f_{\thetahat,(\thetahat-\thetahat^*),\alpha}  \bigg| \\
    &= O_p\bigPar{ \sqrt\frac{{\log r_n}}{r_n} } +o_p \bigPar{  \frac{\log r_n}{n^{\frac{\gamma-2}{2\gamma}}} },
\end{align*}
where the last equality holds by Markov's inequality.

 \end{proof}

\subsection{Auxiliary results for exact estimators}
\label{app:Auxiliary results for exact estimators}

\begin{lem} \label{lem:consitency}
Under Assumptions~\ref{ass:identification}.\ref{ass:identification_global}, \ref{ass:density} and \ref{ass:regressors_instruments},  $ \sup_{\tau \in [\varepsilon,1-\varepsilon]}\| \thetahat - \thetaTrue\|= o_p\bigPar{1}$, where $\thetahat=\thetaLp$ for any $p\in [1,\infty]$ or $\thetahat=\thetaQR$.

\end{lem} 
\begin{proof}
We give the proof for $\thetahat=\thetaLp$. Uniform consistency for the case of $\thetahat=\thetaQR$ was established by \citet[][Theorem 3]{angrist2006quantile}.

By Assumption~\ref{ass:identification}.\ref{ass:identification_global}, 
\begin{equation*}
    \arg\min_{\theta\in\Theta} \|\gt(\theta)\|_p= \thetaTrue.
\end{equation*}
Assumptions~\ref{ass:density} and \ref{ass:regressors_instruments} imply 
that the function class
\begin{align*}
\left\{ (w,y,z) \mapsto z(1\{y\le w'\theta\}-\tau),~ \tau\in[\varepsilon,1-\varepsilon], ~  \theta\in\Theta\right\}
\end{align*}
is Donsker (compare with the function class $\mathcal{F}_n$ in the proof of Lemma \ref{lem:stochasticHolder}) and thus Glivenko-Cantelli, and hence
\begin{equation*}
    \sup_{\theta\in\Theta,\tau \in [\varepsilon,1-\varepsilon]}|\gthat(\theta)-\gt(\theta)| = \sup_{\theta\in\Theta,\tau \in [\varepsilon,1-\varepsilon]}| (\emp-\expect) Z_i(1\{Y_i\le W'\theta\}-\tau)|\asTo 0.
\end{equation*}
By the argmin theorem \citep[e.g., Theorem 2.1 in][]{newey1994large} applied to $ Q_n(\theta,\tau) \bydef \|\gthat(\theta)\|_p$, we get $ \sup_{\tau \in [\varepsilon,1-\varepsilon]}\| \thetaLp- \thetaTrue\|= o_p\bigPar{1}$.  
 \end{proof}
 
\begin{lem}\label{lem:sample_moments_QR} 
Under Assumptions~\ref{ass:identification}--\ref{ass:regressors_instruments}, for any exact QR estimator $\thetaQR$ as defined in equation \eqref{eq:QR_check_function_minimization}, we have
\begin{align}
    \sup_{\tau \in [\varepsilon,1-\varepsilon]} \|\thetaQR - \thetaTrue \| &=   O_p\bigPar{\frac{1}{\sqrt{ n} }}, \label{eq:firstOrderEquivalenceQR} \\
    \sup_{\tau \in [\varepsilon,1-\varepsilon]} \|\gthat(\thetaQR )\|_p &=  O_p    \bigPar{\frac{k}{ n } }.\notag
\end{align}
\end{lem}

\begin{proof}
Equation \eqref{eq:firstOrderEquivalenceQR} follows from Theorem 3 in \citet{angrist2006quantile}.

The exact QR estimators yield exact zeros of the subgradient
\begin{equation*}
    \frac{1}{n}\sum_{i=1}^{n}  (\tau-h(Y_i- W_i^\prime\theta )) W_i, 
\end{equation*}
where the multi-valued function $h(u)$ is defined as $ 1\{u<0\}$ for  $u\neq 0$ and $h(0)\bydef [0,1]$ for $u=0$.
The subgradient function differs from sample moment functions by the fraction of observations with $Y_i= W_i^\prime \thetaQR $.
Note that under Assumption \ref{ass:density}, $Y_i$ has density with respect to the Lebesgue measure conditional on $W_i$. Thus the observations $(Y_i,W_i)$ are in \emph{general position} with probability 1 \citep[see Definition 2.1 and the subsequent discussion in][]{koenker2005quantile}.
Because the observations are in general position with probability 1, there are at most $k$ terms like that, and so $\sup_{\tau \in [\varepsilon,1-\varepsilon]} \|\gthat(\thetaQR )\|_p=O_p\bigPar{ {k}/{n}}$.
\end{proof}

\begin{lem}\label{lem:thetaLp_sqrt_consistency}
Under Assumptions~\ref{ass:identification}, \ref{ass:density}, and \ref{ass:regressors_instruments}, for any estimator $\thetahat=\thetaLp$ that minimizes $\|\gthat(\theta)\|_p$, we have
\begin{align}
   \sup_{\tau \in [\varepsilon,1-\varepsilon]} \|\thetahat - \thetaTrue \| &=   O_p\bigPar{\frac{1}{\sqrt{ n} }},\label{eq:A6_1}\\
    \sup_{\tau \in [\varepsilon,1-\varepsilon]} \|\thetahat - \hat\xi_\tau \| &=   o_p\bigPar{\frac{1}{\sqrt{ n} }}, \label{eq:A6_2} 
    \\
      \sup_{\tau \in [\varepsilon,1-\varepsilon]} \|\gthat(\thetahat)\|_p &= o_p\bigPar{\frac{1}{\sqrt{n}}},\label{eq:A6_3}
\end{align}
where $\hat\xi_\tau$ is defined in equation \eqref{eq:definitionOfTheta_1}.
\end{lem}
\begin{proof}
The proof proceeds in four steps. 

\noindent \textbf{Step 1.} Under the assumptions of the lemma, the empirical process $B_n^\circ(\theta)$ is Donsker (see proof of Lemma \ref{lem:stochasticHolder} above) and thus   asymptotically stochastically equicontinuous \citep[see discussion in][Section 2.1.2; also Theorem 1.5.7 and Problem 2.1.5 in the same book]{van1996weak}. 

\smallskip

\noindent \textbf{Step 2.} 
 By definition, $\sqrt{n}(\hat\xi_\tau-\thetaTrue)$  can be written as
\begin{align}
\sqrt{n} (\hat\xi_\tau -\thetaTrue)  &=    - G^{-1}_\tau \sqrt{n} \bigBra{ \tau(\expect Z- \emp Z)+\frac{1}{\sqrt{n}}  B^\circ_n(\thetaTrue) },    
\end{align} 
where $\thetaTrue$ and $G_\tau \bydef\partial_{\theta} \gt(\thetaTrue)$ are well-defined by Assumption \ref{ass:identification} .
This class of functions indexed by $\tau$ is Donsker by similar arguments as in Lemma \ref{lem:stochasticHolder} (note that the parameter $\tau$ enters this class only through $\thetaTrue$ and through the linear term $\tau Z_i$).
The Donsker property implies  
\begin{equation}
     \sup_{\tau \in [\varepsilon, 1-\varepsilon]}\|   \hat\xi_\tau -\thetaTrue \|= O_p\bigPar{\frac{1}{\sqrt{n}}}.\label{eq:FCLTxi}
\end{equation}
By Lemma~\ref{lem:genericStochasticExpansion} applied to $\hat\xi_\tau$,
\begin{align*}
   \gthat(\hat\xi_\tau ) &=   \frac{1}{\sqrt{n}} B^\circ_n(\thetaTrue) + \tau(\expect Z-\emp{Z}) + \frac{1}{\sqrt{n}}  B_n(\hat\xi_\tau) \notag \\
    &+g( \thetaTrue )+   G(\thetaTrue) (\hat\xi_\tau-\thetaTrue)+   \frac{1 }{2} (   \hat\xi_\tau-\thetaTrue)^\prime \partial_\theta G(\thetaTrue) (\hat\xi_\tau-\thetaTrue) + O_p\bigPar{n^{-\frac{3}{2}}}.
    \end{align*}
    Then  after substituting  the first equation in \eqref{eq:FCLTxi} into the term $ G(\thetaTrue) (\hat\xi_\tau-\thetaTrue)$, we have
    \begin{align*}
   \gthat(\hat\xi_\tau ) = \frac{1}{\sqrt{n}}  B_n(\hat\xi_\tau) + O_p\bigPar{\frac{1}{n}} .
\end{align*}
So by Step 1, $\gthat(\hat\xi_\tau)=O_p\bigPar{n^{-\frac{1}{2}}}$.

Since $\thetaLp$ is defined as the estimator attaining the minimal $p$-norm, 
\begin{equation*}
    \sup_{\tau \in [\varepsilon,1-\varepsilon]} \|\gthat(\thetaLp )\|_p \leq \sup_{\tau \in [\varepsilon,1-\varepsilon]} \|\gthat(\hat\xi_\tau )\|_p = O_p\bigPar{ n^{-\frac{1}{2}}}.
\end{equation*}

\smallskip

\noindent \textbf{Step 3.} 
 Consider $\hat \xi^{(2)}_{\tau}\bydef \hat\xi_\tau-G^{-1} { B_n( \hat\xi_\tau)}/{\sqrt{n}}$.
By equation \eqref{eq:FCLTxi}, $\hat\xi_\tau$ is uniformly consistent, and hence $\hat \xi^{(2)}_{\tau}= \hat\xi_\tau+o_p\bigPar{ {1}/{\sqrt{n}}}$ uniformly in $\tau\in [\varepsilon,1-\varepsilon]$ since $B_n$ is stochastically equicontinuous (Step 1). 
Then by the stochastic equicontinuity of $B^\circ_n$ (Step 1) and  Lemma~\ref{lem:genericStochasticExpansion} applied to $\hat\xi^{(2)}_{\tau}$, uniformly in $\tau \in [\varepsilon,1-\varepsilon]$,
\begin{equation*}
   \gthat(\hat\xi^{(2)}_{\tau}) = \frac{B_n^\circ(\hat\xi^{(2)}_{\tau})-B^\circ_n(\hat\xi_\tau)}{\sqrt{n}}  +o_p\bigPar{\frac{1}{\sqrt{n}}}=o_p\bigPar{\frac{1}{\sqrt{n}}}.
\end{equation*}
This implies 
\begin{equation*}
    \sup_{\tau\in[\varepsilon,1-\varepsilon]} \|\gthat(\thetaLp )\|_p\le \sup_{\tau\in[\varepsilon,1-\varepsilon]}  \|\gthat(\hat\xi^{(2)}_\tau)\|_p= o_p\bigPar{ n^{-\frac{1}{2}}}.
\end{equation*}
which establishes \eqref{eq:A6_3}.

\noindent \textbf{Step 4.}
By Lemma~\ref{lem:consitency}, $ \sup_{\tau \in [\varepsilon,1-\varepsilon]}\| \hat\theta_{\tau,p} -  \thetaTrue  \|= O_p\bigPar{{r^{-1}_n}}$ for some $r_n\to\infty$.
By Lemma~\ref{lem:genericStochasticExpansion} and Steps 1 and 2, $\thetaLp$ satisfies
\begin{align}
    &   G(\thetaTrue) (\thetaLp-\thetaTrue)+   \frac{1 }{2} (   \thetaLp-\thetaTrue)^\prime \partial_\theta G(\thetaTrue) (\thetaLp-\thetaTrue)\notag \\
   &=\gthat(\thetaLp ) - \frac{1}{\sqrt{n}} B^\circ_n(\thetaTrue) - \tau(\expect Z-\emp{Z}) - \frac{1}{\sqrt{n}}  B_n(\thetaLp)  + O_p\bigPar{\frac{1}{r_n^3}}\notag \\
   &= O_p\bigPar{\frac{1}{\sqrt{n}}} +O_p\bigPar{\frac{1}{r_n^3}},  
\end{align}
uniformly in $\tau\in[\varepsilon,1-\varepsilon]$.

By Assumption~\ref{ass:identification}.\ref{ass:identification_full_rank}, we can multiply the last equation by $G^{-1}(\thetaTrue)$ and obtain
\begin{align*}
    \thetaLp-\thetaTrue + O_p\bigPar{\frac{1}{r_n^2}} &= O_p\bigPar{\frac{1}{\sqrt{n}}} +O_p\bigPar{\frac{1}{r_n^3}},
\end{align*}
which implies we can take $r_n= \sqrt{n}  $  by a fixed point argument, which is discussed in detail in Step 3 of the proof of Lemma~\ref{lem:sample_moments_thetaLp} below. This implies \eqref{eq:A6_1}.

By uniform consistency of $\thetaLp$ and Step 1,
\begin{equation*}
  \sup_{\tau \in [\varepsilon,1-\varepsilon]}\| B_n(\thetaLp)\|=\sup_{\tau \in [\varepsilon,1-\varepsilon]}\|B_n^\circ(\thetaLp)-B_n^\circ(\thetaTrue)\| =o_p(1). 
\end{equation*}

Lemma~\ref{lem:genericStochasticExpansion} applied to $\thetaLp$ yields
\begin{align*}
    \thetaLp &=  \hat\xi_\tau  -  G^{-1}(\thetaTrue) \frac{1 }{2} (\thetaLp-\thetaTrue)^\prime \partial_\theta G(\thetaTrue) (\thetaLp-\thetaTrue)+ G^{-1}(\thetaTrue)\gthat(\thetaLp ) \\
    &- \frac{1}{\sqrt{n}}  G^{-1}(\thetaTrue)B_n(\thetaLp)  + O_p\bigPar{n^{-3/2}}.
    \end{align*}
    The term $ G^{-1}(\thetaTrue) 2^{-1} (\thetaLp-\thetaTrue)^\prime \partial_\theta G(\thetaTrue) (\thetaLp-\thetaTrue)$ is $O_p(n^{-1})$. The term $G^{-1}(\thetaTrue)\gthat(\thetaLp )$ is $o_p(n^{-1/2})$ by Step 3. The term $n^{-1/2} G^{-1}(\thetaTrue)B_n(\thetaLp)$ is $o_p(n^{-1/2})$ by stochastic equicontinuity of $B_n$ (Step 1) and uniform consistency of $\thetaLp$.
    Therefore,
\begin{align*}
    \thetaLp &= \hat\xi_\tau + o_p\bigPar{ n^{-\frac{1}{2}}}
\end{align*}
 uniformly in $\tau\in [\varepsilon,1-\varepsilon]$. This proves \eqref{eq:A6_2}.
\end{proof}

The results of the previous lemma can be further refined.

\begin{lem}\label{lem:sample_moments_thetaLp} 
Under Assumptions~\ref{ass:identification}--\ref{ass:regressors_instruments}, for any estimator $\thetaLp$ that minimizes $\|\gthat(\theta)\|_p$, we have
\begin{align*}
       \sup_{\tau \in [\varepsilon,1-\varepsilon]} \|\gthat(\thetaLp )\|_p&=  O_p    \bigPar{\frac{\log n}{ n^{1-\frac{2}{\gamma}} } }.
\end{align*}
\end{lem}

\begin{proof} 
The proof proceeds in four steps.

\smallskip

\noindent \textbf{Step 1.}   By Lemma~\ref{lem:stochasticHolder} and \ref{lem:thetaLp_sqrt_consistency} applied to $\thetaLp$ and $\thetaTrue$, 
\begin{equation*}
    B_n(\thetaLp) = B_n(\thetaLp) -  B_n(\thetaTrue)=   O_p\bigPar{ \sqrt\frac{{\log \sqrt{n}}}{\sqrt{n}} } +o_p \bigPar{  \frac{\log  n }{n^{\frac{\gamma-2}{2\gamma}}} } \quad \text{uniformly in } \tau\in[\varepsilon,1-\varepsilon].
    \end{equation*}
Since $\frac{\gamma-2}{2\gamma}\geq\frac{1}{3}$ for $\gamma\geq6$, 
\begin{equation*}
     B_n(\thetaLp) = O_p\bigPar{ \frac{\sqrt{\log n}}{n^{1/4}}  }\quad \text{uniformly in } \tau\in[\varepsilon,1-\varepsilon].
\end{equation*}

\smallskip

\noindent \textbf{Step 2.} Consider the estimator
\begin{align*}
\hat{\xi}_\tau^{(2)}  &\bydef \hat\xi_\tau   -   \frac{G^{-1}(\thetaTrue)B_n(\thetaLp)}{\sqrt{n}},
\end{align*}  
where, by Step 1, $ {G^{-1}(\thetaTrue)B_n(\thetaLp)}/{\sqrt{n}}=O_p\bigPar{ {n^{-3/4}\sqrt{\log n}}  }$.

By Lemma~\ref{lem:genericStochasticExpansion}, we get
\begin{align}
   \gthat(\hat{\xi}_\tau^{(2)} ) &=   \frac{1}{\sqrt{n}} B^\circ_n(\thetaTrue) + (\tau \expect Z-\tau\emp Z) + \frac{1}{\sqrt{n}} B_n(\hat{\xi}_\tau^{(2)})\notag\\
    &+ G(\thetaTrue) (\hat{\xi}_\tau^{(2)}-\thetaTrue) +   (\hat{\xi}_\tau^{(2)}-\thetaTrue)^\prime  \frac{\partial G(\thetaTrue)}{\partial \theta}(\hat{\xi}_\tau^{(2)}-\thetaTrue) + O_p\bigPar{\frac{1}{n^{3/2}} }.\label{eq:stochExpansionTheta2} 
\end{align}
Then, by definition of $\hat{\xi}_\tau^{(2)}$, 
\begin{align}
   \gthat(\hat{\xi}_\tau^{(2)} ) =& \frac{ B_n(\hat{\xi}_\tau^{(2)} )-B_n(\thetaLp)  }{\sqrt{n}}   + (\hat{\xi}_\tau^{(2)}-\thetaTrue)^\prime  \frac{\partial G(\thetaTrue)}{\partial \theta}(\hat{\xi}_\tau^{(2)}-\thetaTrue)    +O_p\bigPar{\frac{1}{n^{3/2}}} \label{eq:gthetatwo1}.
\end{align}

Define $r_n$ to be a sequence satisfying  $ \sup_{\tau \in [\varepsilon,1-\varepsilon]} \|\thetaLp - \hat{\xi}_\tau^{(2)}\| = O_p(r_n^{-1})$  (Lemma~\ref{lem:thetaLp_sqrt_consistency} implies uniform consistency of $\thetaLp$ and that  $r_n$ can be taken to be at least $\sqrt n$).

By Lemma~\ref{lem:stochasticHolder},  
\begin{equation*}
     B_n(\hat{\xi}_\tau^{(2)} )-B_n(\thetaLp)  =  O_p\bigPar{ \sqrt\frac{{\log r_n}}{r_n} } + o_p\bigPar{\frac{\log r_n}{n^{\frac{\gamma-2}{2\gamma}}}},
\end{equation*}

Then \eqref{eq:gthetatwo1} becomes
\begin{align}
   \gthat( \hat{\xi}_\tau^{(2)} ) =  O_p\bigPar{ \frac{\sqrt{\log n}}{\sqrt{n r_n}} } + o_p\bigPar{\frac{\log n}{ {n^{1-\frac{1}{\gamma}}}}},\label{eq:gtheta2}
\end{align}
where we replaced $\log r_n$ with the faster growing sequence $\log \sqrt{n} = O(\log n)$.

By Lemma~\ref{lem:genericStochasticExpansion} applied to $\thetaLp$ and the definition of $\hat\xi_\tau$,
\begin{align*}
 \thetaLp   &= \hat\xi_\tau   + G^{-1}(\thetaTrue)\gthat(\thetaLp)   -   \frac{G^{-1}(\thetaTrue)B_n(\thetaLp)}{\sqrt{n}}\notag \\
 &-   G^{-1}(\thetaTrue)(\thetaLp-\thetaTrue)^\prime  \frac{\partial G(\thetaTrue)}{\partial \theta}(\thetaLp-\thetaTrue) + O_p\bigPar{\frac{1}{n^{3/2}} }.
\end{align*}
So by \eqref{eq:stochExpansionTheta2} and the definition of $\hat{\xi}_\tau^{(2)}$, we get 
\begin{equation*}
     \thetaLp - \hat{\xi}_\tau^{(2)}= G^{-1}(\thetaTrue)\gthat(\thetaLp)+O_p(n^{-1}),
\end{equation*}
which implies we can take $r_n^{-1}$ as the rate of $\gthat(\thetaLp) = O_p(r_n^{-1})$.

\smallskip

\noindent \textbf{Step 3.} By definition of $\thetaLp$,  $\sup_{\tau \in [\varepsilon,1-\varepsilon]}\|\gthat({\thetaLp} )\|_p \leq   \sup_{\tau \in [\varepsilon,1-\varepsilon]}\|\gthat({\hat{\xi}_\tau^{(2)}} )\|_p$.
Then from   \eqref{eq:gtheta2}, we obtain
 \begin{equation*}
   \sup_{\tau \in [\varepsilon,1-\varepsilon]}\|\gthat({\thetaLp} )\|_p \leq  \sup_{\tau \in [\varepsilon,1-\varepsilon]} \|\gthat({\hat{\xi}_\tau^{(2)}} )\|_p=   O_p\bigPar{ \frac{\sqrt{\log n}}{\sqrt{n r_n}} } + o_p\bigPar{\frac{\log n}{ {n^{1-\frac{1}{\gamma}}}}} .
 \end{equation*}

On the right-hand side of this inequality, suppose that the first term dominates the second term. Then we have
\begin{align*}
(r_n^{-1})^{\frac{1}{2} } &=   O\bigPar{\frac{\sqrt{\log n}} { n^{\frac{1}{2}-\frac{1}{\gamma}} } },
 \end{align*}
 or, equivalently,
 \begin{align*}
r_n^{-1} &= O\bigPar{\frac{\log n}{ n^{1-\frac{2}{\gamma}} } } .
 \end{align*}
 By Step 2, it implies the statement of the lemma.

Suppose, instead, that the second term dominates the first term. 
Then by Step 2, $r_n^{-1} = O \bigPar{n^{-1+\frac{1}{\gamma}} \log n} $. 
The statement of the lemma follows.

\end{proof}
 

\subsection{Proof of Theorem~\ref{thm:stochasticExpansion}}

The proof of Theorem \ref{thm:stochasticExpansion} summarizes the results of auxiliary lemmas in Appendices \ref{app:Auxiliary results for generic IVQR estimators} and \ref{app:Auxiliary results for exact estimators}.

\smallskip

\noindent \textbf{Step 1 (Uniform consistency).}  The estimators under consideration are uniformly consistent, $\sup_{\tau \in [\varepsilon,1-\varepsilon]}\|\thetahat-\thetaTrue\|=O_p\bigPar{{1}/{\sqrt{n}}}$. Specifically, the QR estimator is analyzed in Lemma \ref{lem:sample_moments_QR}; the exact IVQR estimator is analyzed in Lemma \ref{lem:thetaLp_sqrt_consistency}.

\smallskip

\noindent \textbf{Step 2 (Generic stochastic expansion).} By Step 1, we can apply Lemma~\ref{lem:genericStochasticExpansion} with $r_n=\sqrt{n}$ to obtain
\begin{equation*}
\thetahat - G^{-1}\gthat(\thetahat)  = \hat\xi_\tau- 
     G^{-1} \bigBra{ \frac{B_n(\thetahat)}{\sqrt{n}}  + \frac{1 }{2} (  \thetahat-\thetaTrue)^\prime \partial_\theta G(\thetaTrue) (\thetahat-\thetaTrue) } + O_p\bigPar{\frac{1}{n^{3/2}}}.
\end{equation*}

\smallskip

\noindent \textbf{Step 3 (Bounds on remainder in asymptotic linear expansion).} Now we can use the remaining lemmas to bound the orders of the terms in the expansion.
The result in equation \eqref{eq:Bn_supbound} follows from Lemma~\ref{lem:stochasticHolder} with $\thetahat^\ast=\thetaTrue$ and $r_n=\sqrt{n}$.
The first equation in \eqref{eq:ghat_supbound} is stated in Lemma~\ref{lem:sample_moments_QR}.
Similarly, Lemma~\ref{lem:sample_moments_thetaLp} yields the second equation in \eqref{eq:ghat_supbound}.
As a result, we have
\begin{equation}
    \sup_{\tau \in [\varepsilon,1-\varepsilon]}\|\thetahat- \hat\xi_\tau\| =   O_p\bigPar{\frac{\sqrt{\log n}}{n^{3/4}}},
\end{equation}
which is a uniform Bahadur-Kiefer expansion for both QR and exact IVQR estimators. 

\smallskip

\noindent \textbf{Step 4 (Analysis of quadratic term).}
The empirical process $\sqrt{n}( \hat\xi_\tau-\thetaTrue)$ indexed by $\tau\in[\varepsilon,1-\varepsilon]$ is Donsker (see  Step 2 of Lemma \ref{lem:thetaLp_sqrt_consistency}), which implies \eqref{eq:DonskerLinTerm}, i.e.,
\begin{equation}
    \sup_{\tau \in [\varepsilon,1-\varepsilon]}\| \hat\xi_\tau-\thetaTrue\| =   O_p\bigPar{\frac{ 1}{n^{1/2}}}.
\end{equation}
Then by Step 3, we have, uniformly in $\tau\in [\varepsilon,1-\varepsilon]$,
\begin{equation*}
   (   \thetahat-\thetaTrue)^\prime \partial_\theta G(\thetaTrue) (\thetahat-\thetaTrue) = (   \hat\xi_\tau-\thetaTrue)^\prime \partial_\theta G(\thetaTrue) (\hat\xi_\tau-\thetaTrue)  + O_p\bigPar{\frac{\sqrt{\log n}}{n^{5/4}}}.
\end{equation*}
Hence, the expansion in Step 2 becomes  
\begin{equation*}
\thetahat  = \hat\xi_\tau+ 
     G^{-1} \bigBra{\gthat(\thetahat) - \frac{B_n(\thetahat)}{\sqrt{n}}  - \frac{1 }{2} (   \hat\xi_\tau-\thetaTrue)^\prime \partial_\theta G(\thetaTrue)  (\hat\xi_\tau-\thetaTrue) } +  R_{n,\tau} ,
\end{equation*}
with $  \sup_{\tau \in [\varepsilon,1-\varepsilon]}\|R_{n,\tau}\|= O_p\bigPar{\frac{\sqrt{\log n}}{n^{5/4}}}$.

\smallskip

  \qed

\section{Second-order bias correction, proofs}

\subsection{Auxiliary results}

\begin{lem}\label{lem:existence_bias_Bn}
Consider any random sequence $\hat\theta\in \Theta$ a.s.
Under Assumption \ref{ass:regressors_instruments}, the following expectations exist,
\begin{align}
     \expect  \|\gthat(\hat\theta)\| = O(1),\label{eq:existence_bias_Bn1}\\
     \expect \|B^\circ_n(\hat\theta)\| = O(1).\label{eq:existence_bias_Bn2}
\end{align}
 
\end{lem}

\begin{proof}
By the triangular inequality,
    \begin{equation}
         \|\gthat(\hat\theta)\|  \leq \frac{1}{n}\sum_{i=1}^n \|Z_i\|(1+\tau)
    \end{equation}
Therefore, $\expect  \|\gthat(\hat\theta)\|  \leq   2 \expect\|Z\|.$ Since the right-hand side does not depend on $n$, equation \eqref{eq:existence_bias_Bn1} holds.

By definition, we have
\begin{equation}
     B^\circ_n(\theta)    \bydef  \sqrt{n}  (\emp 1\{Y\leq W^{\prime} \theta \} Z  - \expect 1\{Y\leq W^{\prime} \theta \} Z ). 
\end{equation}
We can bound $\expect \|B^\circ_n(\hat\theta)\| $ using a maximal inequality for an appropriately chosen empirical process.
Consider the functions 
\begin{align*}
    &f_{\theta,\alpha}:\, (y,w,z) \mapsto 1\{ y- w^\prime \theta \le 0\}\alpha^\prime z,
\end{align*}
and the corresponding function class
\begin{align*}
    \mathcal{F} &= \bigg\{ f_{\theta,\alpha}: \,\, \theta\in\Theta, \|\alpha\|= 1 \bigg\}.
\end{align*}
Note that $\eG_n f_{\theta,e_j}=B^\circ_n(\hat\theta)^\prime e_j$, where $e_j\bydef(0,\dots 0, 1, 0,\dots 0)^\prime$ with 1 in $j$-th position. By Assumption \ref{ass:regressors_instruments},  $\mathcal{F}$ admits an envelope $F(y,w,z)\equiv \|z\|$. 

We use the maximal inequality in \citet{ota2019quantile} to establish \eqref{eq:existence_bias_Bn2}. To do so, we verify the three conditions of this lemma. First,  
\begin{align*}
    \expect f_{\theta,\alpha}^2(Y,W,Z) &= \E\left[ 1\big\{ Y- W^\prime \theta \le 0\big\} \alpha'Z \right]^2 \le \E \|Z\|^2.
\end{align*}
Therefore, the  variance parameter  of the process is
\begin{equation*}
    \sigma_n^2 \bydef \sup_{f\in\mathcal{F}} \expect f^2(Y,W,Z) \le \E \|Z\|^2.
\end{equation*}

Second, using Lemma 2 in \cite{ota2019quantile} we have $$\expect\max_{1\leq i\leq n } F^2(Y_i,W_i,Z_i)=\expect \max_{1\leq i\leq n } \|Z_i\|^2 = o(n^{2/\gamma}). $$

Third, because the function class $\mathcal{F}$ is a VC class with the   envelope $\|Z\|$, there exist constants $A$ and $V$ independent of $n$ such that the standard entropy bound
\begin{align*}
    \sup_Q N\left( \mathcal{F},\|\cdot\|_{Q,2},  \eta \|Z\|_{Q,2} \right) \le (A / \eta)^V \text{ for all } \eta\in(0,1]
\end{align*}
holds \citep[e.g.,][Section 2.6]{van1996weak}. Here the supremum is taken with respect to all finitely discrete measures $Q$ and $\|\cdot\|_{Q,2}$ is the $L^2(Q)$ norm.

Finally, applying Lemma 1 in \citet{ota2019quantile}, we obtain
\begin{eqnarray}
    &&\expect \sup_{\theta\in\Theta, \|\alpha\|=1}\|\eG_n f_{\theta, \alpha}\|  \lesssim  \sqrt{V \sigma_n^2 \log (A m/ \sigma_n ) } +  \frac{V \sqrt{\expect \max_{1\leq i\leq n } \|Z_i\|^2 }}{\sqrt{n}} \log (A m/\sigma_n)\notag \\
    &&=  O\bigPar{ 1 } +o \bigPar{  \frac{1}{n^{\frac{\gamma-2}{2\gamma}}} }   ,\label{eq:useful_maximal_inequality2}
\end{eqnarray}

It follows that
\begin{equation*}
   \|B^\circ_n(\hat\theta)\| \le \sup_{\theta\in\Theta, \|\alpha\|=1}\|\eG_n f_{\theta, \alpha}\|
\end{equation*}
which implies
\begin{equation*}
   \expect\|B^\circ_n(\hat\theta)\| \le \expect \sup_{\theta\in\Theta, \|\alpha\|=1}\|\eG_n f_{\theta, \alpha}\| = O\bigPar{ 1 } . 
\end{equation*}
 
\end{proof}



     






\begin{lem}\label{lem: mean squared convergence}
Consider $\hat\theta_n$  such that  $\hat\theta_n\in\Theta$ and  $\hat\theta_n-\thetaTrue  = o_p(1)$. Then $\expect\|\hat\theta_n-\thetaTrue\|^q=o(1)$ for any $q>0$.
\end{lem}

\begin{proof}
   Notice that  $$\|\hat\theta_n-\thetaTrue\|^q\leq (\max_{\theta\in\Theta} \|\theta-\thetaTrue\|)^q \le \text{diam}(\Theta)^q. $$
Hence the sequence $\|\hat\theta_n-\thetaTrue\|^q$ is uniformly integrable.
By Proposition 4.12 from \cite{kallenberg2006foundations}, 
$\expect\|\hat\theta_n-\thetaTrue\|^q\ =o(1)$.

\end{proof}

\begin{lem}\label{lem:auxiliary_bias}
Consider $\hat\theta=\thetaLp$ defined in \eqref{eq:GMM_Lp} for some $p\in [1,\infty]$ or $\hat\theta=\thetaQR$, where $\tau \in (0,1)$.
Under Assumptions~\ref{ass:identification}, \ref{ass:density}, and \ref{ass:regressors_instruments}, $ B_n(\hat\theta)=B^1_n + B^2_n $, where the two components satisfy
\begin{equation*}
    \expect \frac{1}{\sqrt{n}} B^1_n (\hat\theta)=  \expect \bigPar{ \frac{\gthat(\hat\theta) + \gthat^*(-\hat\theta) }{2}} +  \frac{1}{n}\kappa(\tau),
\end{equation*}
with
\begin{equation*}
    \kappa(\tau)  \bydef  \expect  \bigPar{\tau -\frac{1}{2} } f_{ \varepsilon} (0 |W, Z) Z W^{\prime} G^{-1}Z ,   
\end{equation*}
while $B^2_n  =O_p\bigPar{n^{-3/4}\sqrt{\log  n}}$ and $n^{-1/2}  B^2_n  $ is uniformly integrable.
\end{lem}

\begin{proof} The proof proceeds in six steps.

\smallskip

\noindent\textbf{Step 1.} By Lemma \ref{lem:existence_bias_Bn}, $\expect B_n(\hat\theta) $  exists. 
Note that
\begin{align}
    \frac{1}{\sqrt{n}}\expect B_n(\hat\theta) &= \frac{1}{\sqrt{n}}\expect  \bigPar{ B^\circ_n(\hat\theta) - B^\circ_n(\thetaTrue) }\notag\\
    & = \expect  \bigPar{  1\{Y\leq 
        W^{\prime} \hat{\theta} \}    Z  } - \expect   g^{\circ}(\hat\theta). \label{eq:step1_two_terms}
\end{align}

Theorem~\ref{thm:stochasticExpansion} implies 
\begin{equation}
       \hat\theta = \thetaTrue -\frac{1}{n}G^{-1}\sum_{i=1}^n\bigPar{1\{Y_i\le W_i^\prime \thetaTrue\}-\tau}Z_i + \tilde{R}_n ,\label{eq:weakBahadurKiefer}
\end{equation}
where $\tilde{R}_n =O_p\bigPar{n^{-3/4}\sqrt{\log{n}}}$.
Since by construction, $ \hat\theta $ is restricted to a compact set $\Theta$, it is bounded. 
The term $\bigPar{1\{Y_i\le W_i^\prime \thetaTrue\}-\tau}Z_i$ has bounded moments up to order $\gamma$ by Assumption~\ref{ass:regressors_instruments}. 
As a result, the remainder term $\tilde R_n$ has bounded moments up to order $\gamma$.

\smallskip

\noindent\textbf{Step 2.} Define $\hat\varepsilon_i\bydef Y_i- W_i^\prime \hat\theta$ and split the first term in equation \eqref{eq:step1_two_terms} as follows:
\begin{equation}
     \expect 1\{Y_i\leq W_i^\prime \hat\theta\}Z_i = \expect 1\{\hat\varepsilon_i=0\}Z_i  + \expect 1\{\hat\varepsilon_i <  0 \}Z_i. \label{eq:influenceSplit}
\end{equation}

We can use \eqref{eq:weakBahadurKiefer} to isolate an influence of observation $i$, $ \lambda_i \bydef -  W_i ^\prime G^{-1} Z_i (1\{Y_i\le W_i^\prime\thetaTrue\}-\tau)$. 
Without loss of generality for i.i.d.\ data, we consider $i=1$.
The indicator $1\{\hat\varepsilon_1 <  0 \}$ can be rewritten as
$   1 \left\{ Y_1< W_1^\prime\hat \theta_{-1}+ n^{-1}\lambda_1 \right\},$ 
where 
\begin{equation*}
       \hat\theta_{-1}\bydef \hat\theta + \frac{1}{n}G^{-1}\bigPar{1\{Y_1\le W_1^\prime \thetaTrue\}-\tau}Z_1 = \thetaTrue -\frac{1}{n}G^{-1}\sum_{j= 2}^n\bigPar{1\{Y_j\le W_j^\prime \thetaTrue\}-\tau}Z_j +  \tilde{R}_n
\end{equation*} 
is equal to $\hat\theta$ without the linear influence of the observation $i=1$.

Then, using Taylor's theorem (justified below equation \eqref{eq:conditional_density_bounds}), 
\begin{align}
 &\expect \left[ Z_1 P\bigPar{Y_1< W_1^\prime \hat\theta_{-1}+ \frac{1}{n}\lambda_1 |1\{Y_1\le W_1'\thetaTrue\},Z_1,W_1} \right] \notag\\
 &= \expect \left[ Z_1 P\bigPar{Y_1< W_1^\prime \hat\theta_{-1}|1\{Y_1\le W_1'\thetaTrue\},Z_1,W_1} \right]\notag \\
 &+\expect  \left[ \frac{1}{n} Z_1\lambda_1  f_{Y_1}(W_1^\prime \hat\theta_{-1} |1\{Y_1\le W_1'\thetaTrue\}, Z_1,W_1) \right] + \frac{1}{n^2}\psi^1_n, \notag 
\end{align}
where by the mean value theorem $\psi^1_n = O(1)$.
 
Note also that the first term in the Taylor expansion can be rewritten as
\begin{equation*}
     \expect \left[Z_1 P(Y_1< W_1^\prime \hat\theta_{-1}|1\{Y_1\le W_1'\thetaTrue\},Z_1,W_1)\right] = \expect \left[ Z_1 1\{Y_1< W_1^\prime \hat\theta_{-1}\} \right]
     = \expect \left[ Z_1 P(Y_1< W_1^\prime \hat\theta_{-1}|Z_1, W_1) \right].
\end{equation*}

The following argument justifies the use of Taylor's theorem here. The function $f_{Y_1}(y |\hat\theta_{-1},\lambda_1,W_1,Z_1)$  is measurable as a limit of measurable functions (increments of conditional CDF).
Therefore, for any non-negative measurable function $\phi(W_1,Z_1)$ with finite expectation, the integral \\
$ \expect\left[ \phi(W_1,Z_1) f_{Y_1}(y |\hat\theta_{-1},\lambda_1,W_1,Z_1) \right]$ exists (but may take infinite values).
By the law of iterated expectations,
\begin{align*}
    \expect \left[ \phi(W_1,Z_1) f_{Y_1}(y |\hat\theta_{-1},\lambda_1,W_1,Z_1) \right] = \expect  \left[\phi(W_1,Z_1) f_{Y_1}(y |W_1,Z_1) \right] 
\end{align*}
(see Step 5 below for a detailed justification based on Fubini-Tonelli theorem).
By Assumption~\ref{ass:density}, $f_{Y_1}(y |W_1,Z_1)$ is uniformly bounded and
\begin{align}
 \expect \left[ \phi(W_1,Z_1) f_{Y_1}(y |W_1,Z_1) \right]\le \bar f \cdot \expect \left[ \phi(W_1,Z_1) \right] < \infty .\label{eq:conditional_density_bounds}
\end{align} 
The same is true for the derivative of the density $\partial f_{Y_1}$ in place of $f_{Y_1}$, by Assumption \ref{ass:density}.
Therefore, $P (  f_{Y_1}(y |\hat\theta_{-1},\lambda_1,W_1,Z_1)=\infty )=0$ and $P (  \partial f_{Y_1}(y |\hat\theta_{-1},\lambda_1,W_1,Z_1)=\infty )=0$, which justifies the Taylor expansion of the expectations of the conditional PDF  above.
By this property (a.s.\ smoothness of $f_{Y_1}( y |\hat\theta_{-1},\lambda_1,Z_1, W_1 )$) and equation \eqref{eq:weakBahadurKiefer},
\begin{align*}
    \expect \left[ Z_1 \lambda_1 f_{Y_1}(W_1^\prime \hat\theta_{-1} | \hat\theta_{-1},W_1,\lambda_1, Z_1 ) \right] & =  \expect \left[ Z_1 \lambda_1f_{Y_1}(W_1^\prime \thetaTrue |W_1,Z_1,\lambda_1) \right] \\
    &+ \expect \bigPar{ Z_1 \lambda_1 W_1^\prime (\hat\theta_{-1}-\thetaTrue) \partial f_{Y_1}(\xi |W_1,Z_1,\lambda_1)},
    \end{align*} 
where   $\xi$  is some random variable that takes values between  $W_1^\prime  \hat\theta_{-1}$ and $W_1^\prime\thetaTrue$.
By the boundedness  of $\hat\theta_{-1} \in \Theta$, Assumption \ref{ass:regressors_instruments}, the bound on the derivative of the density in Assumption \ref{ass:density}, and the fact that  $\hat\theta_{-1}=\thetaTrue +  O_p\bigPar{{1}/{\sqrt{n}}}$,  these expectations exist and the second term denoted as
$$\psi^2_n \bydef   Z_1 \lambda_1 W_1^\prime (\hat\theta_{-1}-\thetaTrue) \partial f_{Y_1}(\xi |W_1,Z_1,\lambda_1)  $$ 
is of order  $  O_p\bigPar{{1}/{\sqrt{n}}}$. 

By the definition of $\lambda_1$, the first term can be rewritten as
\begin{align*}
    \expect \left[ Z_1 \lambda_1 f_{\varepsilon_1}(0 |W_1,Z_1,\lambda_1)\right]  = & -\expect \left[ Z_1  W_1 ^\prime G^{-1} Z_1  1\{\varepsilon_i \le  0\}   f_{\varepsilon_1}(0 |W_1,Z_1,\lambda_1) \right]\\
    &+ \expect \left[  Z_1  W_1 ^\prime G^{-1} Z_1 \tau f_{\varepsilon_1}(0 |W_1,Z_1) \right].  
\end{align*} 

Finally, \eqref{eq:influenceSplit} becomes 
\begin{align}
&\expect \left( 1\{Y_i\leq W_i^\prime \hat\theta\}Z_i - \frac{1}{n} \psi^2_n\right) =\notag
 \expect \left[ Z_1 P(Y_1< W_1^\prime \hat\theta_{-1}|Z_1,W_1) \right] \\
 &+ \frac{\tau}{n}\expect\left[ f_{\varepsilon_1}(0|W_1,Z_1)Z_1 W_1^\prime G^{-1} Z_1 \right] +  \expect \left[ 1\{\hat\varepsilon_1 =  0 \}Z_1 \right] +\frac{1}{n}\Xi_\tau 
 +  \psi_n^1,
 \label{eq:TaylorStrictInequality2} 
\end{align} 
where the term $\Xi_\tau \bydef -\expect \left[ Z_1  W_1 ^\prime G^{-1} Z_1  1\{\varepsilon_1 \le  0\}   f_{\varepsilon_1}(0 |W_1,Z_1,\lambda_1) \right]$ and $ 
\frac{1}{n} \psi^2_n = O_p\bigPar{ n^{-3/2} }.
$ 
 
\smallskip

\noindent\textbf{Step 3.} Now consider $\expect g^{\circ}(\hat\theta)$, the second term in \eqref{eq:step1_two_terms}. 
Let $(Y_{n+1},W_{n+1},Z_{n+1})$ be a copy of $(Y,W,Z)$, which is independent of the sample $\{Y_i,W_i,Z_i\}_{i=1}^n$.
Also, define 
\begin{equation*}
\lambda_{n+1,1} \bydef - \frac{1}{n} W_{n+1}^\prime G^{-1} Z_1 (1\{Y_1\le W_1^\prime\thetaTrue\}-\tau),    
\end{equation*}
which satisfies $\expect \lambda_{n+1,1} =0$.
Then
\begin{align}
     \expect\left(g^{\circ}(\hat\theta) -\frac{1}{n}\psi^4_n \right) &= \expect  (1\{Y_{n+1}\leq 
        W_{n+1}^{\prime} \hat{\theta} \}    Z_{n+1} -\frac{1}{n}\psi^4_n)\notag \\
    &=  \expect  \left(P\{Y_{n+1}\leq 
        W_{n+1}^{\prime} \hat\theta_{-1}  - \frac{1}{n}\lambda_{n+1,1}|  W_{n+1},Z_{n+1}\}    Z_{n+1} -\frac{1}{n}\psi^4_n\right) \notag \\
   & =    \expect \left( P\{Y_{n+1}< 
        W_{n+1}^{\prime} \hat \theta_{-1}|  W_{n+1},Z_{n+1}\}    Z_{n+1}\right) \notag \\
       &+ \frac{1}{n}\expect \left( Z_{n+1}\lambda_{n+1,1}  f_{Y_{n+1}}(W_{n+1}^\prime \thetaTrue |1\{Y_{1}\le W_{1}'\thetaTrue\}, Z_{n+1},W_{n+1}) \right) \label{eq:LamdaNplus1}
       \\
        & +  \frac{1}{n^2}\psi_n^3 ,\notag
\end{align} 
where $\psi_n^3=O(1)$ by the mean value theorem and 
$$
\psi^4_n \bydef    Z_{n+1} \lambda_{n+1} W_{n+1}^\prime (\hat\theta_{-1}-\thetaTrue) \partial f_{Y_{n+1}}(\xi |W_{n+1},Z_{n+1},\lambda_{n+1}).
$$
The rate  $\psi^4_n  = O_p(n^{-1/2})$ is derived by an argument similar to the one below equation \eqref{eq:conditional_density_bounds}. 
Note that the term in line \eqref{eq:LamdaNplus1} is equal to zero since $\expect \bigPar{\lambda_{n+1,1} | Y_{n+1},W_{n+1},Z_{n+1}}=0$ by the i.i.d.\ data assumption.
Combining this equality with \eqref{eq:TaylorStrictInequality2} yields
\begin{align}
      & \expect  \bigPar{  1\{Y_1\leq 
        W_1^{\prime} \hat{\theta} \}    Z_1  } -\expect g^\circ(\hat\theta)\notag-\frac{1}{n}\expect (\psi^2_n- \psi^4_n)-   \frac{\psi_n^1-\psi_n^3}{n^2}\\ 
       & = \expect \left[ Z_1 P\{Y_{1}< 
        W_{1}^{\prime} \hat \theta_{-1}|  W_{1},Z_{1}\}  \right] -\expect \left[ Z_{n+1} P\{Y_{n+1}< 
        W_{n+1}^{\prime} \hat \theta_{-1}|  W_{n+1},Z_{n+1}\} \right]\notag \\
       &+   \frac{\tau}{n}\expect\left[ Z_1 W_1^\prime G^{-1} Z_1   f_{\varepsilon_{1}}(0|W_1,Z_1) \right] +  \expect 1\{\hat\varepsilon_1 =  0 \}Z_1 +\frac{1}{n}\Xi_\tau. \label{eq:intermediateBiasFormula}
\end{align}

\smallskip

\noindent\textbf{Step 4.} Let us simplify the first two terms of equation \eqref{eq:intermediateBiasFormula}.
Define  
\begin{equation*}
     \hat\zeta_{-1} \bydef - \frac{1}{n}  G^{-1}\sum_{j=2}^n  \bigPar{1\{\varepsilon_j\le 0 \}-\tau}Z_j,
\end{equation*}
so that $ \hat\zeta_{-1}$ has zero mean and is independent of $Y_1$ and $\hat\theta_{-1}=\thetaTrue + \hat\zeta_{-1}+\tilde R_n$.

Denote $\hat\xi_1   \bydef Y_1 - W_1^\prime  \hat\zeta_{-1}$. 
Apply Taylor's theorem (as in Step 2) to obtain
\begin{align}
     &\expect \left[ Z_1 P\{Y_{1}< 
        W_{1}^{\prime} \hat \theta_{-1}|  W_{1},Z_{1}, W_1^\prime \tilde R_n\} \right]\notag \\
    &= \expect \left[ Z_1 P\{\hat\xi_1 < 
        W_{1}^{\prime} (\thetaTrue + \tilde R_{n})|  W_{1},Z_{1}, W_1^\prime \tilde R_n\} \right] \notag \\
     & =  \expect \left[ Z_1 P(\hat\xi_1 < W_{1}^\prime \thetaTrue | W_1,Z_1) \right] + \expect \left[ Z_1 W_1^\prime  \tilde R_n  f_{\hat\xi_1}(W_{1}^\prime\thetaTrue| W_1,Z_1, \tilde R_n)\right] \notag\\
     &+  \frac{1}{2}\expect \left[ Z_1 \tilde R_n^\prime W_1 \partial f_{\hat\xi_1}(\eta| W_1,Z_1, \tilde R_n)W_1^\prime \tilde R_n \right]  , \label{eq:strictCDFTaylorExpansion}
\end{align}
where $\eta$ is a random scalar that takes values between  $W_{1}^\prime\thetaTrue$ and $W_{1}^\prime \tilde R_n$.
By Step 1, $\tilde R_n =O_p\bigPar{ n^{-3/4}\sqrt{\log  n}}$ has   a finite second moment. Therefore, the last term in \eqref{eq:strictCDFTaylorExpansion} is finite.

For the second term in \eqref{eq:strictCDFTaylorExpansion}, note that
\begin{align*}
     \tilde{R}_n f_{ \hat{\xi}_1}(W_{1}^\prime\thetaTrue| W_1,Z_1,\tilde{R}_n)  &= \tilde{R}_n f_{ \varepsilon_1}(W_1^\prime \hat{\zeta}_{-1} | W_1,Z_1,\tilde{R}_n)\\
 &=\tilde{R}_n f_{ \varepsilon_1}(0 | W_1,Z_1,\tilde{R}_n) +\partial f_{ \varepsilon_1}(\tilde{\eta} | W_1,Z_1,\tilde{R}_n) \tilde{R}_n W_1^\prime\hat{\zeta}_{-1} \\
  &=\tilde{R}_n f_{ \varepsilon_1}(0| W_1,Z_1,\tilde{R}_n) + O_p\bigPar{\frac{\sqrt{\log n}}{n^{5/4}}},
\end{align*}
where $\tilde\eta$ is a random scalar that takes values between $0$ and $W_1'\hat\zeta_{-1}$. The last equality follows since $\tilde{R}_n = O_p\bigPar{ n^{-3/4}\sqrt{\log  n}}$ by Step 1 and $\hat{\zeta}_{-1}= O_p\bigPar{1/{\sqrt{n}}}$.
As before, let us introduce 
\begin{equation*}
    \psi^5_n \bydef  \partial f_{ \varepsilon_1}(\tilde{\eta} | W_1,Z_1,\tilde{R}_n) \tilde{R}_n W_1^\prime \hat{\zeta}_{-1}  +\frac{1}{2}  Z_1 \tilde R_n^\prime W_1 \partial f_{\hat\xi_1}(\tilde\eta| W_1,Z_1, \tilde R_n)W_1^\prime \tilde R_n = O_p\bigPar{\frac{\sqrt{\log n}}{n^{5/4}}}.
\end{equation*}
By the boundedness of $\gamma$-moments of $\tilde R_n $ (Assumption \ref{ass:regressors_instruments}), $\expect \psi^5_n$ exists. 
Hence, \eqref{eq:strictCDFTaylorExpansion} becomes
\begin{equation*}
\expect \left[ Z_1 P(Y_1 - W_1^\prime  \hat\zeta_{-1} < W_{1}^\prime \thetaTrue | W_1,Z_1)\right] + \expect \left[ Z_1 W_{1}^\prime \tilde{R}_n  f_{ \varepsilon_1}(0| W_1,Z_1,\tilde{R}_n)\right]   + \expect \psi^5_n.\label{eq:correlationRn1}
\end{equation*} 

Similarly, using the i.i.d.\ assumption, 
\begin{align*}
&\expect\left[  Z_{n+1} P\{Y_{n+1}< 
        W_{n+1}^{\prime} \hat \theta_{-1}|  W_{n+1},Z_{n+1}\}\right] \\
        & = \expect \left[ Z_{n+1} P(Y_{n+1} - W_{n+1}^\prime \hat\zeta_{-1} < W_{n+1}^\prime \thetaTrue | W_{n+1},Z_{n+1})\right] \\
        &+ \expect \left[ Z_{n+1}  W_{n+1}^\prime \tilde{R}_n  f_{ \varepsilon_{n+1}}( W_{n+1}^\prime \hat\zeta_{-1}| W_{n+1},Z_{n+1},\tilde{R}_n) \right]   + \expect \psi^6_n\\
     & = \expect\left[ Z_1 P(Y_1 - W_1^\prime \hat\zeta_{-1
,n} < W_1^\prime \thetaTrue | W_{1},Z_{1}) \right] +   \expect \left[ f_{ \varepsilon_{1}}(0| W_{1},Z_{1}) Z_{1
}  W_{1}^\prime \right] \cdot \expect \tilde{R}_n  +    \expect  \psi^6_n,
\end{align*}
where
\begin{equation*}
    \psi^6_n \bydef  \partial f_{ \varepsilon_1}(\tilde{\eta} | W_1,Z_1) \tilde{R}_n W_1^\prime \hat{\zeta}_{-1}  +\frac{1}{2}  Z_1 \tilde R_n^\prime W_1 \partial f_{\hat\xi_1}(\tilde\eta| W_1,Z_1)W_1^\prime \tilde R_n = O_p\bigPar{\frac{\sqrt{\log n}}{n^{5/4}}}.
\end{equation*}
To summarize, \eqref{eq:intermediateBiasFormula} becomes
\begin{align}
      & \expect  \bigPar{  1\{Y_1\leq 
        W_1^{\prime} \hat{\theta} \}Z_1} -\expect   g^{\circ}(\hat\theta) - \expect  \bigPar{\frac{\psi^1_n-\psi^3_n}{n^2}+\frac{\psi_n^2-\psi^4_n}{n}+\psi_n^5-\psi^6_n} \notag \\
       & =   \frac{\tau}{n}\expect \left[ Z_1 W_1^\prime G^{-1} Z_1  f_{\varepsilon_1}(0|W_1,Z_1)\right] +  \expect 1\{\hat\varepsilon_1=  0 \}Z_1+\frac{1}{n}\Xi_\tau   \notag\\
       &+ \expect \left[ Z_{1}  W_{1}^\prime (\tilde{R}_n-\expect \tilde{R}_n)  (f_{ \varepsilon_{1}}(0| W_{1},Z_{1},\tilde{R}_n) -f_{ \varepsilon_{1}}(0| W_{1},Z_{1})) \right]. \label{eq:intermediateBiasFormula20}
\end{align}

\smallskip

\noindent\textbf{Step 5.}
Let us study the last term in equation \eqref{eq:intermediateBiasFormula20}.
For any $t\geq0$, consider an auxiliary function
\begin{align*}
\Psi(t) &\bydef\expect \left[ Z_{1}  W_{1}^\prime (\tilde{R}_n-\expect \tilde{R}_n)  (F_{ \varepsilon_{1}}(t| W_{1},Z_{1},\tilde{R}_n) - F_{ \varepsilon_{1}}(t| W_{1},Z_{1})) \right].
\end{align*}
By definition, for every $t\geq 0$,
\begin{equation*}
 \Psi(t)   =    \expect\left[ Z_{1}  W_{1}^\prime (\tilde{R}_n-\expect \tilde{R}_n)   1\{0<\varepsilon_{1}\le t\} \right]    -    \expect\left[ Z_{1}  W_{1}^\prime (\tilde{R}_n-\expect \tilde{R}_n)      1\{0<\varepsilon_{1}\le t\}\right] =0 .
\end{equation*}
By the existence of the corresponding conditional PDF (possibly taking infinite values),  
$$\Psi(t) = \expect \left[ Z_{1}  W_{1}^\prime (\tilde{R}_n-\expect \tilde{R}_n)  \int _{-\infty}^t(f_{ \varepsilon_{1}}(e| W_{1},Z_{1},\tilde{R}_n) - f_{ \varepsilon_{1}}(e| W_{1},Z_{1}))\,de \right] .$$
 By the Fubini-Tonelli theorem for product measures, we can exchange the order of integration,
 $$\Psi(t) = \int _{-\infty}^t \expect \left[ Z_{1}  W_{1}^\prime (\tilde{R}_n-\expect \tilde{R}_n)  (f_{ \varepsilon_{1}}(e| W_{1},Z_{1},\tilde{R}_n) - f_{ \varepsilon_{1}}(e| W_{1},Z_{1}))\right] \,de.$$
 Hence, by the main theorem of  calculus, for all $e\geq 0$,
 $$ \frac{\partial  \Psi(e)}{\partial e}  = \expect \left[ Z_{1}  W_{1}^\prime (\tilde{R}_n-\expect \tilde{R}_n)  (f_{ \varepsilon_{1}}(e| W_{1},Z_{1},\tilde{R}_n) -f_{ \varepsilon_{1}}(e| W_{1},Z_{1})) \right].
$$ 
Since the function $\Psi(t) \equiv 0$, we have
\begin{align*}
     \frac{\partial  \Psi(0)}{\partial e} = \expect \left[ Z_{1}  W_{1}^\prime (\tilde{R}_n-\expect \tilde{R}_n)  (f_{ \varepsilon_{1}}(0| W_{1},Z_{1},R_n) -f_{ \varepsilon_{1}}(0| W_{1},Z_{1}))\right] = 0.
\end{align*}
Therefore, equation \eqref{eq:intermediateBiasFormula20} becomes
\begin{align}
      & \expect  \bigPar{  1\{Y_1\leq 
        W_1^{\prime} \hat{\theta} \}Z_1} -\expect   g^{\circ}(\hat\theta) - \expect  \psi_n \notag \\ 
       & =   \frac{\tau}{n}\expect\left[ Z_1 W_1^\prime G^{-1} Z_1  f_{\varepsilon_1}(0|W_1,Z_1)\right] +  \expect 1\{\hat\varepsilon_1=  0 \}Z_1+\frac{1}{n}\Xi_\tau, \label{eq:intermediateBiasFormula2}
\end{align}
where $$\psi_n\bydef\bigPar{\frac{\psi^1_n-\psi^3_n}{n^2}+\frac{\psi_n^2-\psi^4_n}{n}+\psi_n^5-\psi^6_n}=O_p\bigPar{\frac{\sqrt{\log n}}{n^{5/4}}}.$$

\smallskip

\noindent\textbf{Step 6.} Let us simplify the second and the third terms in equation \eqref{eq:intermediateBiasFormula2}. The latter can be rewritten as
\footnotesize
\begin{align}
       &  \expect \bigPar{  1\{Y_1\leq 
        W_i^{\prime} \hat{\theta} \}    Z_1  } - \expect \bigPar{  1\{Y_{n+1}\leq 
        W_{n+1}^{\prime} \hat{\theta} \}    Z_{n+1}  }    \label{eq:SampleMoment1}\\
       & = \expect 1\{\hat\varepsilon_1 =  0 \}Z_1 - \expect  \bigPar{  1\{Y_1\ge 
        W_i^{\prime} \hat{\theta} \}    Z_1  } +\expect   \bigPar{ 1\{Y_{n+1}\ge
        W_{n+1}^{\prime} \hat{\theta} \}    Z_{n+1}  }\notag\\
       & =\expect 1\{\hat\varepsilon_1 =  0 \}Z_1- \expect  \bigPar{  1\{-Y_1\le 
        W_i^{\prime} (- \hat{\theta}) \} -(1-\tau) }    Z_1 +\expect   \bigPar{ 1\{-Y_{n+1}\le
        W_{n+1}^{\prime}(- \hat{\theta}) \}  -(1-\tau)}   Z_{n+1} \label{eq:SampleMoment2} \\
        & = \expect 1\{\hat\varepsilon_1 =  0 \}Z_1 - \bigg [  \frac{1}{n}\expect\left[ Z_1 W_i^\prime G^{-1} Z_1 (1-\tau) f_\varepsilon(0|W_1,Z_1)\right] + \frac{1}{n}\Xi^*_\tau + \expect 1\{-\hat\varepsilon_1 =  0 \}Z_1 \bigg] +   \expect \psi^*_n \notag\\
        & =    \bigg [  \frac{1}{n}\expect\left[ Z_1 W_1^\prime G^{-1} Z_1 ( \tau-1) f_\varepsilon(0|W_1,Z_1)\right] - \frac{1}{n}\Xi^*_\tau     \bigg] +   \expect \psi^*_n, \label{eq:step5_final_step_1st_argument}
\end{align}
\normalsize
where $\Xi^*_\tau \bydef -\expect \left[ Z_1  W_1 ^\prime G^{-1} Z_1  1\{-\varepsilon_1 \le  0\}   f_{-\varepsilon_1}(0 |W_1,Z_1,1\{-\varepsilon_1\le 0\}) \right]$, $\psi^*_n$ is the analog of $\psi_n$ corresponding to the moment condition for $-Y_i$, $-\theta$, and $(1-\tau)$, and the last equality uses \eqref{eq:intermediateBiasFormula2}. 
Notice that it follows from equations \eqref{eq:SampleMoment1} and \eqref{eq:SampleMoment2} that
\begin{equation}
 \expect 1\{\hat\varepsilon_1 =  0 \}Z_1 = \expect \bigPar{ \hat g(\hat\theta) + \hat g^*(-\hat\theta)   }.  \label{eq:E-eps-Z}
\end{equation}
Using the definition of $f_{\varepsilon_1}(0|W_1,Z_1,1\{\varepsilon_1\le 0\})$ and Fubini-Tonelli theorem as in Step 5,
\begin{align*}
    \Xi_\tau &=-\expect \left[ Z_1  W_1 ^\prime G^{-1} Z_1  1\{\varepsilon_1 \le  0\}   f_{\varepsilon_1}(0 |W_1,Z_1,1\{\varepsilon_1 \le  0\}) \right]\\
     &=-\lim_{t\downarrow 0}\expect \left[ Z_1  W_1 ^\prime G^{-1} Z_1  1\{\varepsilon_1 \le  0\}  \frac{1\{\varepsilon_1 \le  0\}-1\{\varepsilon_1 \le   -t\}}{t} \right]\\
     &=-\lim_{t\downarrow 0}\expect \left[ Z_1  W_1 ^\prime G^{-1} Z_1   \frac{1\{\varepsilon_1 \le  0\}-1\{\varepsilon_1 \le   -t\}}{t} \right]\\
     &=-\expect \left[ Z_1  W_1 ^\prime G^{-1} Z_1     f_{\varepsilon_1}(0 |W_1,Z_1) \right] .
\end{align*}
Since $\varepsilon_1$ has conditional density by Assumption~\ref{ass:density}, the same argument can be applied to show that $\Xi^*_\tau = \Xi_\tau$.
Hence, equations \eqref{eq:intermediateBiasFormula2} and \eqref{eq:step5_final_step_1st_argument} imply
\begin{align*}
    &  \frac{1}{n} \Xi_\tau = -\frac{1}{2} \bigPar{ \expect \bigPar{ \hat g(\hat\theta) + \hat g^*(-\hat\theta)   }  +    \frac{1}{n}\expect\left[  Z_1 W_1^\prime G^{-1} Z_1  f_{\varepsilon_1}(0|W_1,Z_1) \right]  }  +    \frac{1}{2}(\expect \psi_n -\expect \psi^*_n ).
\end{align*}
Finally, equations \eqref{eq:intermediateBiasFormula2} and \eqref{eq:E-eps-Z} yield
\begin{align*}
        \expect \frac{1}{\sqrt{n}} (B_n(\hat\theta) - B^2_n)  
         &=    \frac{1}{n}\expect\left[ Z_1 W_1^\prime G^{-1} Z_1 \bigPar{\tau - \frac{1}{2} } f_\varepsilon(0|W_1,Z_1)\right]  +  \expect \bigPar{ \frac{\hat g(\hat\theta)+ \hat g^*(-\hat\theta) }{2} },
\end{align*}
where $  n^{-1/2}  B^2_n \bydef \frac{1}{2}(  \psi_n +  \psi^*_n) $.
By construction, $B^2_n=O_p\bigPar{ n^{-3/4}\sqrt{\log n}} $ and $ n^{-1/2}  B^2_n $ is uniformly integrable as the sum of uniformly integrable components.
The proof is complete.
 \end{proof}

\begin{lem}\label{thm: olver}
 Suppose that a function $f(x)$ is four times continuously differentiable in a neighborhood of x.
 Then for sufficiently small $h\in\R$,
 \begin{align*}
     \partial_{x}f(x)&=\frac{f(x+h)-f(x-h)}{2h}+O(h^2),\\
     \partial_{x,x} f(x)&=\frac{f(x+h  ) - 2f(x) + f(x-h )}{  h^2}+O(h^2).
 \end{align*}
  
\end{lem} 

\begin{proof}
    See  Chapter 5 in \citet{olver2014introduction} and   p.884 in \citet{abramowitz1972handbook}.
\end{proof}

 \subsection{Proofs of main results on bias correction}

\begin{proof}[Proof of Theorem~\ref{thm:bias}]

By Theorem~\ref{thm:stochasticExpansion},
\begin{equation*}
\thetahat  = \hat\xi_\tau+ 
     G^{-1} \bigBra{\gthat(\thetahat) - \frac{B_n(\thetahat)}{\sqrt{n}}  - \frac{1 }{2} (   \hat\xi_\tau-\thetaTrue)^\prime \partial_\theta G(\thetaTrue) (\hat\xi_\tau-\thetaTrue) } + R_{n,\tau},
\end{equation*}
where $R_{n,\tau}=O_p\bigPar{ n^{-5/4}\sqrt{\log  n}}$, and $\|R_{n,\tau}\|$ is uniformly integrable by Lemmas \ref{lem:existence_bias_Bn} and \ref{lem: mean squared convergence}.
Lemma~\ref{lem:auxiliary_bias} implies 
\begin{equation*}
   \frac{1}{\sqrt{n}}   \expect B^1_n(\thetahat)=  \expect \bigPar{ \frac{\hat g(\hat\theta) + \hat g^*(-\hat\theta) }{2}} +  \frac{1}{n}\kappa(\tau) .\label{eq:implication_lemma_disaster}
\end{equation*}
For correctly specified models, $\expect \hat \xi_\tau = \thetaTrue $ and, for each component $j$, we have
\begin{align*}
    (   \hat\xi_\tau-\thetaTrue)^\prime \partial_\theta G_j(\thetaTrue) (\hat\xi_\tau-\thetaTrue)  
    &=     \expect   \gthat^\prime(\thetaTrue)  ( G^{-1})^\prime \partial_\theta G_j(\thetaTrue) G^{-1}  \gthat(\thetaTrue)   \\
    &= \frac{1}{n}Q_j^\prime vec( \Omega ).
\end{align*}
By definition of $\bias(\thetahat)$, we can ignore the terms $R_{n,\tau}$ and $B_n^2$. The statement of the theorem follows.
 \end{proof}

 \begin{proof}[Proof of Theorem~\ref{thm:biascorrection}]

Theorem \ref{thm:stochasticExpansion} implies the following asymptotic expansion for the bias-corrected estimator,
\begin{align}
  \hat\theta_{bc}&=\hat\xi_\tau+ 
     G^{-1} \bigBra{\gthat(\thetahat) - \frac{B_n(\thetahat)}{\sqrt{n}}  - \frac{1 }{2} (   \hat\xi_\tau-\thetaTrue)^\prime \partial_\theta G(\thetaTrue) (\hat\xi_\tau-\thetaTrue) }\label{eq:feasible_bc_1}\\
     &- \expect {G}^{-1}\frac{\bigPar{\gthat(\thetahat)-\gthat^*(-\thetahat)}}{2} + \frac{1}{n}   G^{-1} \bigBra{  \kappa_{\tau }  + \frac{1}{2}   Q^\prime vec(  \Omega ) }\label{eq:feasible_bc_2} \\
     &-  (\hat{G}^{-1}- {G}^{-1})\frac{\bigPar{\gthat(\thetahat)-\gthat^*(-\thetahat)}}{2} + \frac{1}{n} (\hat G^{-1}-{G}^{-1}) \bigBra{   \hat \kappa_{\tau }  + \frac{1}{2} \hat Q^\prime vec( \hat \Omega ) }\label{eq:feasible_bc_3}\\
     &+  \frac{1}{n}   G^{-1} \bigBra{   \hat \kappa_{\tau }-   \kappa_{\tau }  + \frac{1}{2} \hat Q^\prime vec( \hat \Omega )-  Q^\prime vec(  \Omega )  } + R_{n,\tau} \label{eq:feasible_bc_4}\\
    &   -   {G}^{-1}\frac{\bigPar{\gthat(\thetahat)-\gthat^*(-\thetahat)}}{2}+\expect {G}^{-1}\frac{\bigPar{\gthat(\thetahat)-\gthat^*(-\thetahat)}}{2}  \label{eq:feasible_bc_5}.
\end{align}
By Theorem \ref{thm:bias}, the expectation of the sum of the terms in \eqref{eq:feasible_bc_1} and \eqref{eq:feasible_bc_2} is zero.
By Theorem \ref{thm:stochasticExpansion}, Assumption \ref{ass:consistent_estimators}, and the Mann-Wald and Delta theorems, the first term  in \eqref{eq:feasible_bc_3}  is \[o_p\bigPar{\frac{1}{n^{1/3}\log n}} O_p\bigPar{\frac{\log n}{n^{1-2/\gamma}}}=o_p\bigPar{n^{-1}},\]
since $\gamma\geq6$.
The same rate $o_p\bigPar{n^{-1}}$ is true for the second term in \eqref{eq:feasible_bc_3} and the terms in \eqref{eq:feasible_bc_4}
The last line, expression \eqref{eq:feasible_bc_5}, has zero mean by Assumption \ref{ass:regressors_instruments}. 
Therefore, $\bias(\hat\theta_{bc})=0$.
 \end{proof}

 \begin{proof}[Proof of Lemma~\ref{thm: finite difference}]
 
 Notice that by the same arguments as in Lemma~\ref{lem:stochasticHolder}, for any $h$,
  \begin{align*}
      &\emp\frac{1\{Y\le W'\thetahat +h\} }{2h}Z_i W_j -\expect \frac{ F_{Y}( W' \theta +h |W,Z)  }{2 h}Z_i W_j \bigg |_{\theta=\thetahat} \\
      &- \left(\emp\frac{1\{Y\le W'\thetahat -h\} }{2h}Z_i W_j -\expect \frac{ F_{Y}( W' \theta -h |W,Z)  }{2 h}Z_i W_j \bigg |_{\theta=\thetahat}\right)= O_p\bigPar{\frac{\sqrt{\log h}}{\sqrt{n h}}}+o_p \bigPar{  \frac{\log h}{n^{5/6} h} }.
 \end{align*}
Then using Lemma~\ref{thm: olver} for $h_n \to 0$ and the Delta theorem, we obtain
 \begin{align*}
    &\emp\frac{1\{Y\le W'\thetahat+ h_{n}\} - 1\{Y\le W'\thetahat- h_{n}\} }{2h_{n}}Z_i W_j \notag \\
    &=   G_{i,j}(\theta )  +O_p\bigPar{\frac{1}{\sqrt{n}}} +O\bigPar{h_n^2} + O_p\bigPar{\frac{\sqrt{\log h_n}}{\sqrt{nh_n}}}+o_p \bigPar{  \frac{\log h_n}{n^{5/6} h_n} }.
 \end{align*}
The overall rate is $\max\left\{n^{-1/2},h_n^2,(nh_n)^{-1/2} \sqrt{\log h_n}, (n^{5/6} h_n)^{-1}\log h_n \right\} $. 
By Lemmas~\ref{lem:stochasticHolder}, \ref{lem:sample_moments_QR}, \ref{lem:sample_moments_thetaLp},  and Assumption \ref{ass:density}, this remainder rate is uniform in $\tau\in[\varepsilon,1-\varepsilon]$.
Ignoring a logarithmic factor, we see that the bandwidth $h_{1,n}\propto {n^{-1/5}} $ delivers an optimal overall remainder rate $O_p\bigPar{  n^{-2/5} \sqrt{\log n}}$ that is uniform in $\tau\in[\varepsilon,1-\varepsilon]$ .

Similarly, by Lemma~\ref{thm: olver}, 
  \begin{align*}
   & \widehat{({\partial_\theta G}_{i,\ell})}_j =   e_i^{\prime}\partial_\theta G_j(\theta)e_\ell +O_p\bigPar{\frac{1}{\sqrt{n}}} +O(h_n^2) + O_p\bigPar{\frac{\sqrt{\log h_n}}{\sqrt{n}\sqrt{h_n^3}}} + o_p \bigPar{  \frac{\log h_n}{n^{5/6} h_n^2} }.
  \end{align*}
 Taking $h_{2,n} \propto n^{-1/7} $, we obtain the optimal remainder rate
   \begin{equation*}
     \widehat{({\partial_\theta G}_{i,\ell})}_j= e_i^{\prime} \partial_\theta G_j(\theta)e_\ell + O_p\bigPar{\frac{\sqrt{\log n}}{  n^{2/7} }}.
 \end{equation*}
 
 Notice that
 \begin{align*}
      \hat Q_j & =  Q_j + O  \bigPar{\max \{\|\hat G-G\|, \|\widehat{({\partial_\theta G}_{i,\ell})}_j- {({\partial_\theta G}_{i,\ell})}_j\|\}   } =  Q_j +O_p\bigPar{\frac{\sqrt{\log n}}{  n^{2/7} }}.
 \end{align*}

By an argument similar to the above,  
  \begin{align*}
   \hat \kappa_{\tau}& =  \bigPar{\tau-\frac{1}{2}} \emp \left[ \frac{1\{Y\le W'\thetahat+ h_{n}\} - 1\{Y\le W'\thetahat- h_{n}\} }{2h_{n}} Z W^{\prime}   G^{-1}Z \right]\\
   &+  \sum_{j=1}^k \bigPar{\tau-\frac{1}{2}} \emp \left[ \frac{1\{Y\le W'\thetahat+ h_{n}\} - 1\{Y\le W'\thetahat- h_{n}\} }{2h_{n}} (e'_jZ_j)  Z W^{\prime}    \right](\hat G^{-1} - G^{-1})e_j\\
   &=\kappa_{\tau} +O_p\bigPar{\frac{1}{\sqrt{n}}}+ O_p\bigPar{\frac{\sqrt{\log h_{n}}}{\sqrt{n h_{n}}}}  + o_p \bigPar{  \frac{\log h_{n}}{n^{5/6} h_{n}} }+O(h_{n}^2)+ O_p\bigPar{ \frac{ \sqrt{\log n}}{n^{2/5}} }.
 \end{align*} 
 where the last term is based on the fact  
 \begin{equation}
     \|\hat G^{-1}-G^{-1}\|\leq \|\hat G^{-1}\|_2  \|G^{-1}\|_2 \|\hat G -G\| .
 \end{equation}
So the overall rate is
 \begin{align*}
     \max\left\{  n^{-1/2}, \,  \frac{\sqrt{\log n}}{n^{2/5}  }, \, \frac{\log h_{n}}{n^{5/6} h_{n}}, \, h_{n}^2, \, \frac{\sqrt{\log h_{n}}}{\sqrt{n h_{n}}}\right \}.
 \end{align*}
 The optimal bandwidth is $h_{3,n} \propto n^{-1/5}$ with
 \begin{equation*}
   \hat \kappa_{\tau} =  \kappa_{\tau}   + O_p \bigPar{  \frac{\sqrt{\log n}}{n^{2/5}} }. 
 \end{equation*}
 By the CLT and the equicontinuity of the relevant sample moment functions implied by Lemma~\ref{lem:stochasticHolder},
 \begin{align*}
     \hat \Omega_{\tau} &\bydef \widehat{Var} [Z(1\{Y\le W' \thetahat\}-\tau)] \\
     &= \emp [(1\{Y\le W' \thetahat\}-\tau)ZZ'] \notag \\
     &- \emp [Z(1\{Y\le W' \thetahat\}-\tau)]\emp [Z'(1\{Y\le W' \thetahat\}-\tau)] \\
     &=\Omega_\tau + O_p\bigPar{ \frac{1}{\sqrt{n}} }.
\end{align*}
\end{proof}

\section{Illustration of approximate bias formula in univariate case}
\label{app: Illustration of approximate bias formula in univariate case}

Suppose we are interested in estimating the $\tau$-quantile of a uniformly distributed outcome variable $Y$.
This is a special case of the general framework with $W=Z=1$, $f_Y(y)=1\{0\leq y\leq1\}$. 

Note that, under the maintained assumptions, the true parameter $\thetaTrue$ has an equivalent alternative definition as a solution to 
\begin{equation*}
 \expect[(1\{- Y\leq W^{\prime}(-\thetaTrue)\}-(1-\tau)) Z ] = 0. \label{eq:unconditionalMoment_symmetrized}
\end{equation*}
As a result, there are two ways of defining an estimator: as a minimizer of 
$|\gthat (\theta)|$ or as a minimizer of $|\gthat^* (-\theta)|$, where
\begin{align*}
    &\gthat (\theta) = \emp (1\{Y\le \theta \}-\tau),\\
    & \gthat^* (-\theta) = \emp 1\{-Y\le -\theta \}-(1-\tau).
\end{align*}
The derivatives of the population moment conditions $\gt(\theta)=\gt^*(-\theta)=0$ are $G=1$, $\partial_\theta G  = 0 $ and $G^*\bydef\partial_\theta \gt^*(-\theta)=-1$, $\partial_\theta G^*  = 0 $, respectively.
In either case, the closure of the argmin set will be $[Y_{(k)},Y_{(k+1)}]$, where $k\bydef \intPart{\tau n}$.
If the fractional part $\{\tau n\} \bydef \tau n-\intPart{\tau n} \leq\frac{1}{2}$, a minimizer of $|\gthat (\theta)|$ ($|\gthat^* (-\theta)|$) is the order statistic $Y_{(k)}$ ($Y_{(k+1)}$, respectively); if $\{\tau n\} \geq\frac{1}{2}$, a minimizer of $|\gthat (\theta)|$ ($|\gthat^* (-\theta)|$) is $Y_{(k+1)}$ ($Y_{(k)}$, respectively). Of course, on the real line $\mathbb{R}^1$, all norms $\|\cdot\|_p,~ p\in [1,\infty]$, coincide with the absolute value $|\cdot|$.  

In this simple example, formula \eqref{eq:biasFormula} yields asymptotic bias expansions
\begin{align}
&\expect Y_{(k)} - \tau = 
\frac{k-\tau n}{n}
+  \frac{1}{n} \bigPar{\frac{1}{2} - \tau} - \frac{1}{2n} +o\bigPar{\frac{1}{n}}= 
-\frac{\{\tau n \}}{n}-\frac{\tau}{n} +o\bigPar{\frac{1}{n}},\label{eq:asy_bias_uniform_1} \\
&\expect Y_{(k+1)} - \tau  =  
\frac{k-\tau n}{n} +  \frac{1}{n} \bigPar{\frac{1}{2} - \tau} + \frac{1}{2n}+ o\bigPar{\frac{1}{n}}=-\frac{\{\tau n \}}{n}+ \frac{1-\tau}{n} +o\bigPar{\frac{1}{n}}. \label{eq:asy_bias_uniform_2}
\end{align}
The exact bias formulas are given by \citep[e.g.,][]{ahsanullah2013introduction}
\begin{align*}
     &\expect Y_{(k)} - \tau = \frac{k }{n+1} -\tau =-\frac{\{\tau n \}}{n+1}- \frac{\tau}{n+1},\\
     &\expect Y_{(k+1)} - \tau = \frac{k +1}{n+1} -\tau = -\frac{\{\tau n \}}{n+1}+ \frac{1-\tau}{n+1}.
\end{align*}
Comparing these formulas with the asymptotic formulas \eqref{eq:asy_bias_uniform_1} and \eqref{eq:asy_bias_uniform_2}, we see that they indeed coincide up to $O\bigPar{ n^{-2} }$.
Figure \ref{fig:exact_asymptotic_uniform_quantile} in the main text illustrates the exact and the second-order bias formula (scaled by $n$) for $n=10$.

\section{Exact QR and IVQR algorithms}\label{app:LP_MILP}

First consider a linear programming (LP) implementation of the QR regression \eqref{eq:QR_check_function_minimization} \citep[][Section 6.2]{koenker2005quantile}:
\begin{align*}
    &\min_{ \theta,r,s } \tau \iotabf'r + (1-\tau)  \iotabf's\\
    s.t.~~&\varepsilon_i=r_i - s_i=Y_i-W_i'\theta, \,\,i=1,\dots,n,\\
    & r_i\geq 0, s_i \geq 0, \,\,i=1,\dots,n.
\end{align*}
Here $\iotabf$ is an $(n\times 1)$ vector of ones. This formulation allows us to apply LP solvers like Gurobi to obtain the exact minimum in  \eqref{eq:QR_check_function_minimization}. 

Next consider the exact estimator for the IVQR case,
 \begin{equation*}
\hat{\theta}_{\tau,1} = \operatorname{argmin}_{\theta\in \Theta} ||\gthat(\theta)||_1.  \label{eq:GMM_Lp_MILP}   
\end{equation*}
The underlying optimization problem can be equivalently reformulated as a mixed integer linear program (MILP) with special ordered set (SOS) constraints,
\begin{align*}
    &\min_{e,\theta,r,s,t} \iotabf't\\
    s.t.~~&\varepsilon_i=r_i - s_i=Y_i-W_i'\theta, \,\,i=1,\dots,n,\\
    &(r_i,e_i)\in SOS_1, \,\,i=1,\dots,n,\\
    &(s_i,1-e_i)\in SOS_1, \,\,i=1,\dots,n,\\
    & r_i\geq 0, s_i \geq 0, \,\,i=1,\dots,n,\\
    & e_i \in \{0,1\} , \,\,i=1,\dots,n,\\
    &-t_l \leq {Z}_l'(e-\tau\iotabf) \leq t_l,\,\, l=1,\dots,d.
\end{align*}
where ${Z}_l$ is an $n\times1$ vector of realizations of instrument $l$.
All constraints except the last one coincide with the ones derived by \citet{chen2018exact} in  Appendix C.1 (we also omit the redundant constraint $r_i+s_i>0$, which is implied by the two $SOS_1$ constraints).
The last constraint ensures that the objective function is the $\ell_1$ norm of the just identifying moment conditions.

\begin{rem}
We also considered the ``big-M'' formulation while performing the Monte Carlo analyses.
The big-M formulation has certain computational advantages, although the arbitrary choice of tuning parameters may result in sub-optimal solutions. 
This problem is more prominent for tail quantiles.
Since the big-M formulation does not guarantee exact solutions, consistent with our theory, the choice of tuning parameters may affect the asymptotic bias.
We prefer the above SOS formulation because it does not depend on tuning parameters as the big-M MILP/MIQP formulations in \citet{chen2018exact} and \citet{zhu2019learning}.\footnote{These papers pick the value of the tuning parameter $M$ as a solution to a linear program that in turn depends on the choice of an arbitrary box around a linear IV estimate. This is problematic if there is a lot of heterogeneity in the coefficients across quantiles. Moreover, in linear models with heavy tailed residuals, the linear IV estimator is not consistent.} \qed
\end{rem}

\section{Stochastic expansion of 1-step corrected IVQR estimators}
\label{app: Stochastic expansion of general quantile regression estimators}

In the main text, we focus in classical QR and exact IVQR estimators.
As shown in the following corollary, the results in Theorem \ref{thm:stochasticExpansion} can be used to obtain a uniform BK expansion for general IVQR estimators after a feasible 1-step correction. 

\begin{cor}\label{lem:BK_generic}
Suppose that Assumptions~\ref{ass:identification}--\ref{ass:regressors_instruments} hold.
Consider any estimator $\thetahat$ such that $ \sup_{\tau \in [\varepsilon,1-\varepsilon]}\| \thetahat - \thetaTrue\|= O_p\bigPar{n^{-1/2}}$.
Then 
\begin{equation*}
  \sup_{\tau \in [\varepsilon,1-\varepsilon]}\| \thetahat  - \hat G^{-1}\gthat(\thetahat) - \hat\xi_\tau \| =   O_p\bigPar{\frac{\sqrt{\log n}}{n^{3/4}}},
\end{equation*} 
where $\hat{G}$ is defined in \eqref{eq:G-hat-ij}.
\end{cor}
\begin{proof}

By Lemma~\ref{lem:genericStochasticExpansion} applied to $\thetahat$, uniformly in $\tau\in [\varepsilon,1-\varepsilon]$,
\begin{equation*}
\thetahat - G^{-1}\gthat(\thetahat)  = \hat\xi_\tau+O_p\bigPar{\frac{\sqrt{\log n}}{n^{3/4}}}.
\end{equation*}

Under the maintained assumptions,  Lemma \ref{thm: finite difference} implies
\begin{equation*}
    \sup_{\tau \in [\varepsilon,1-\varepsilon]}\| \hat G (\thetahat)-G (\thetaTrue) \| = O_p\bigPar{ n^{-2/5}\sqrt{\log n} }.
\end{equation*}
By Lemma \ref{lem:differentiability_g(theta)}, $\partial G(\theta)$ is bounded  uniformly over $\theta \in \Theta$.
Then, by Assumption \ref{ass:identification}.\ref{ass:identification_full_rank} and continuity of the minimal eigenvalue function, the eigenvalues of $G(\theta)$ are bounded away from zero on $\theta\in\Theta$. 
Therefore, the derivative of the inverse matrix function, $F(A)\bydef A^{-1}$, is uniformly bounded over $G (\thetaTrue)$ for $\tau \in [\varepsilon,1-\varepsilon]$.
Hence, by the element-wise Taylor expansion of $F$ at $G(\thetaTrue)$,
\begin{align*}
    \sup_{\tau \in [\varepsilon,1-\varepsilon]}\| \hat G^{-1}(\thetahat)-G^{-1}(\thetaTrue) \| = O \bigPar{\sup_{\tau \in [\varepsilon,1-\varepsilon]}\|   \hat G (\thetahat)-G (\thetaTrue) \|}
    =    O_p\bigPar{ n^{-2/5}\sqrt{\log n} }.
\end{align*}
By Lemma~\ref{lem:stochasticHolder}, \begin{equation}
  \sup_{\tau \in [\varepsilon,1-\varepsilon]}\|\gthat(\thetahat)- \gthat(\thetaTrue) \| =  O_p \bigPar{  \frac{\sqrt{\log n}}{n^{3/4}} }.  \label{eq:sampleMomErr}
 \end{equation}
By Donsker's theorem,
\begin{equation*}
  \sup_{\tau \in [\varepsilon,1-\varepsilon]}\|\gthat(\thetaTrue)\|  = O_p\bigPar{  \frac{1}{n^{1/2}} }  ,
\end{equation*}
so, by the triangular inequality and  \eqref{eq:sampleMomErr},
\begin{equation*}
  \sup_{\tau \in [\varepsilon,1-\varepsilon]}\|\gthat(\thetahat)\|  = O_p\bigPar{  \frac{1}{n^{1/2}} }.
\end{equation*}
Then 
\begin{equation*}
    \sup_{\tau \in [\varepsilon,1-\varepsilon]}\| \hat G^{-1}(\thetahat)\gthat(\thetahat)-G^{-1}(\thetaTrue)\gthat(\thetahat) \| = O_p\bigPar{n^{-1/2}  n^{-2/5}\sqrt{\log n} },
\end{equation*}
which concludes the proof.
\end{proof}

\newpage
\section{Additional figures}\label{sec:additionalFigures}

In this section, we present additional simulation results. To explore the impact of the strength of the instruments, we consider two additional DGPs.

\begin{center}
\begin{tabular}{ l  l  l}
\hline
\hline
 DGP7 (Uniform, endogenous, weaker) &$F_U(u)=\int_{-\infty}^u 1\{t\in[0,1]\}dt$ & $\Sigma_{12}=0.6$, $\Sigma_{13}=0.25$\\
\hline
DGP8 (Uniform, endogenous, stronger) &$F_U(u)=\int_{-\infty}^u 1\{t\in[0,1]\}dt$ & $\Sigma_{12}=0.9$, $\Sigma_{13}=0.25$\\
 \hline
 \hline
\end{tabular}
\end{center}

In DGP7, the instrument is weaker than in DGP1, and in DGP8, the instrument is stronger than in DGP4.

 \begin{figure}[H]
	\caption{Bias (multiplied by $n$) before and after correction, IV strength grows from left to right}

	\begin{center}
	\begin{subfigure}[b]{0.3\textwidth}
		\includegraphics[width=\textwidth]{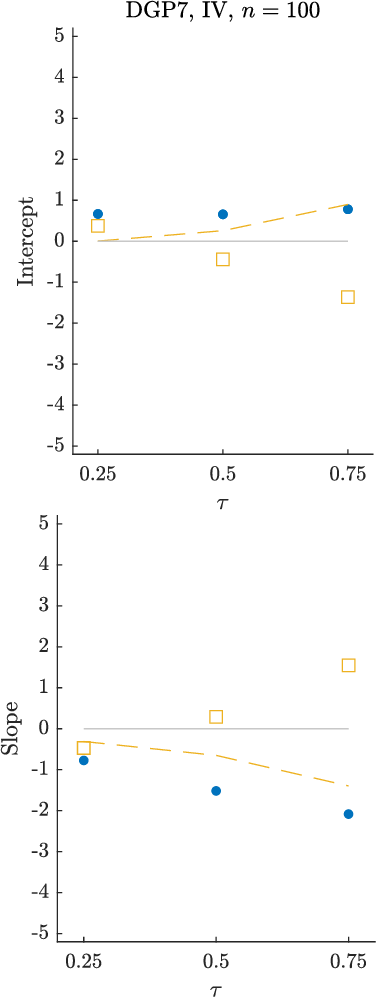}
		
	\end{subfigure}
	\begin{subfigure}[b]{0.3\textwidth}
		\includegraphics[width=\textwidth]{sections/output/uniform_locScaleEND/IV/opt_feasible.eps}
	
	\end{subfigure}
	\begin{subfigure}[b]{0.3\textwidth}
		\includegraphics[width=\textwidth]{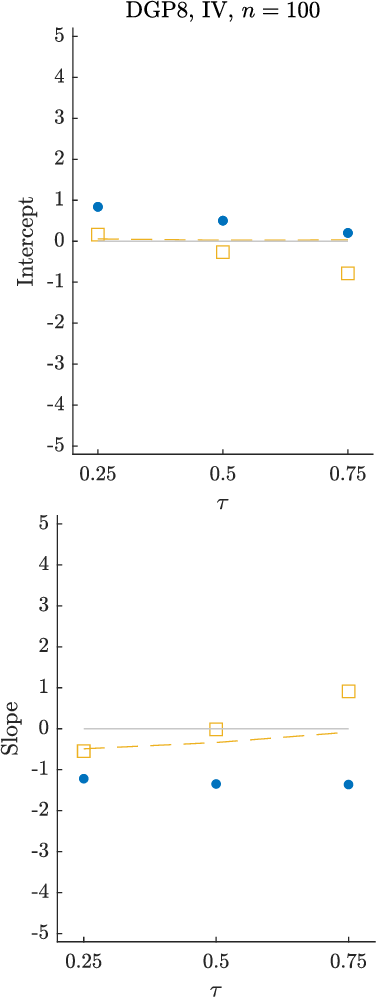}
	
	\end{subfigure}
		\end{center}
	  \footnotesize{\textit{Notes:} The panels display the bias (multiplied by $n$) of the intercept and the slope for IVQR (implemented via the MILP formulation in Appendix \ref{app:LP_MILP}) without bias correction (blue dots), IVQR with feasible bias correction based on the rule-of-thumb bandwidth (gold squares), and IVQR with infeasible bias correction (gold dashed line) for DGP7, DGP4, and DGP8 (ordered in terms of instrument strength). All results are based on 5,000 simulation repetitions. The infeasible bias correction is based on the feasible formula applied to a simulated sample of 10,000,000 observations.}
\label{fig:IV_strength}
 
\end{figure}

 \begin{figure}[H]
	\caption{Bias (multiplied by $n$) before and after correction for DGP1, different sample sizes}

	\begin{center}
	\begin{subfigure}[b]{0.3\textwidth}
		\includegraphics[width=\textwidth]{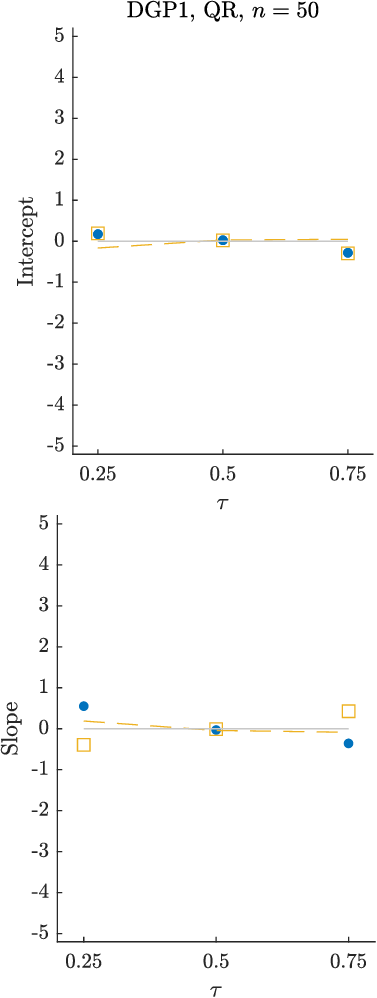}
		
	\end{subfigure}
	\begin{subfigure}[b]{0.3\textwidth}
		\includegraphics[width=\textwidth]{sections/output/uniform_locScaleEX/QR/opt_feasible.eps}
	
	\end{subfigure}
	\begin{subfigure}[b]{0.3\textwidth}
		\includegraphics[width=\textwidth]{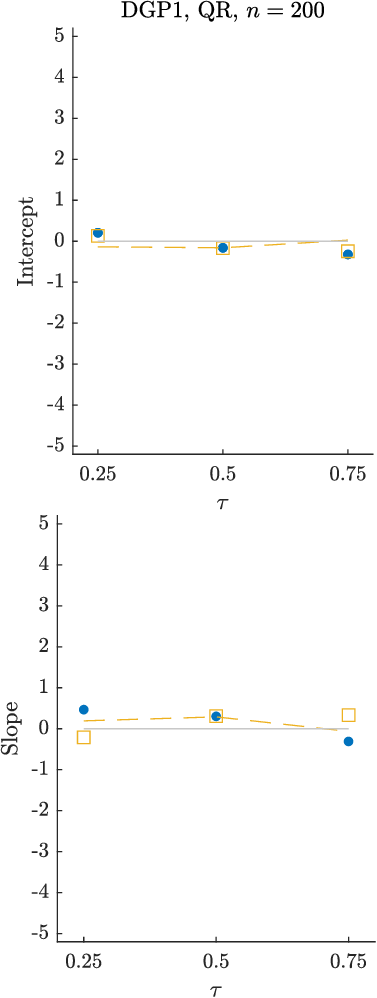}
	
	\end{subfigure}
		\end{center}
	  \footnotesize{\textit{Notes:} The panels display the bias (multiplied by $n$) of the intercept and the slope for classical QR without bias correction (blue dots), QR with feasible bias correction based on the rule-of-thumb bandwidth (gold squares), and QR with infeasible bias correction (gold dashed line) for DGP1 with $n\in \{50,100,200\}$. All results are based on 5,000 simulation repetitions.}
\end{figure}

 \begin{figure}[H]
	\caption{Bias (multiplied by $n$) before and after correction for DGP4, different sample sizes}

	\begin{center}
	\begin{subfigure}[b]{0.3\textwidth}
		\includegraphics[width=\textwidth]{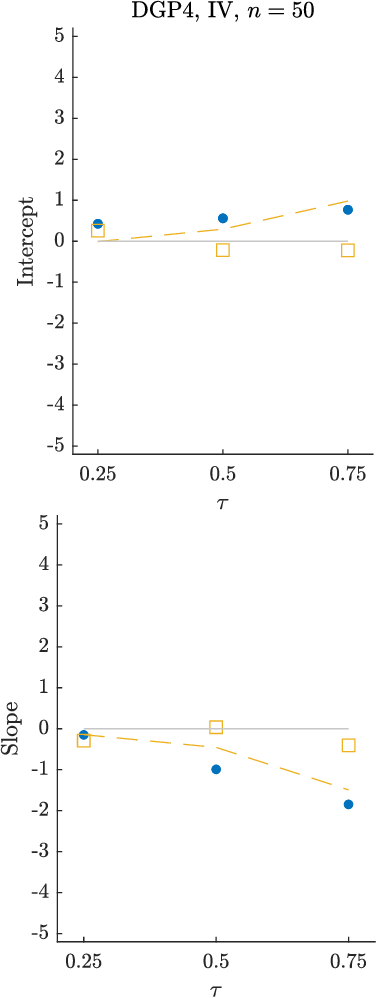}
		
	\end{subfigure}
	\begin{subfigure}[b]{0.3\textwidth}
		\includegraphics[width=\textwidth]{sections/output/uniform_locScaleEND/IV/opt_feasible.eps}
	
	\end{subfigure}
	\begin{subfigure}[b]{0.3\textwidth}
		\includegraphics[width=\textwidth]{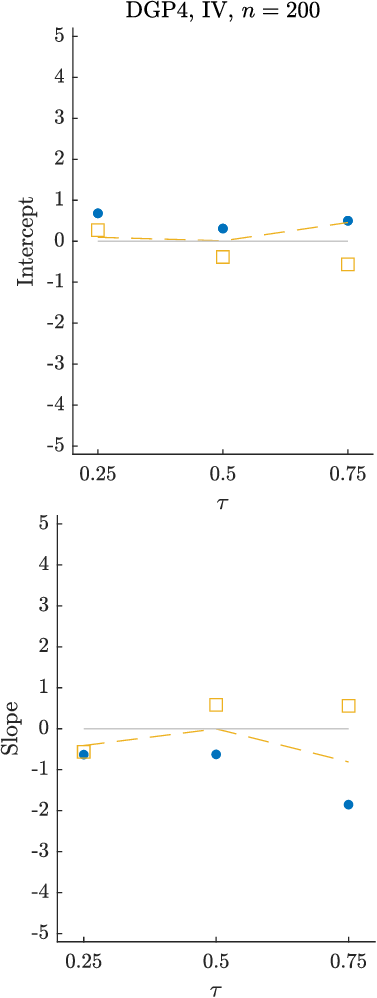}
	
	\end{subfigure}
		\end{center}
	  \footnotesize{\textit{Notes:} The panels display the bias (multiplied by $n$) of the intercept and the slope for IVQR (implemented via the MILP formulation in Appendix \ref{app:LP_MILP}) without bias correction (blue dots), IVQR with feasible bias correction based on the rule-of-thumb bandwidth (gold squares), and IVQR with infeasible bias correction (gold dashed line) for DGP4 with $n\in \{50,100,200\}$. All results are based on 5,000 simulation repetitions. The infeasible bias correction is based on the feasible formula applied to a simulated sample of 10,000,000 observations.}

\end{figure}

\begin{figure}[H]
	\caption{RMSE comparison of raw and bias-corrected estimators }

	\begin{center}
	\begin{subfigure}[b]{0.45\textwidth}
		\includegraphics[width=\textwidth]{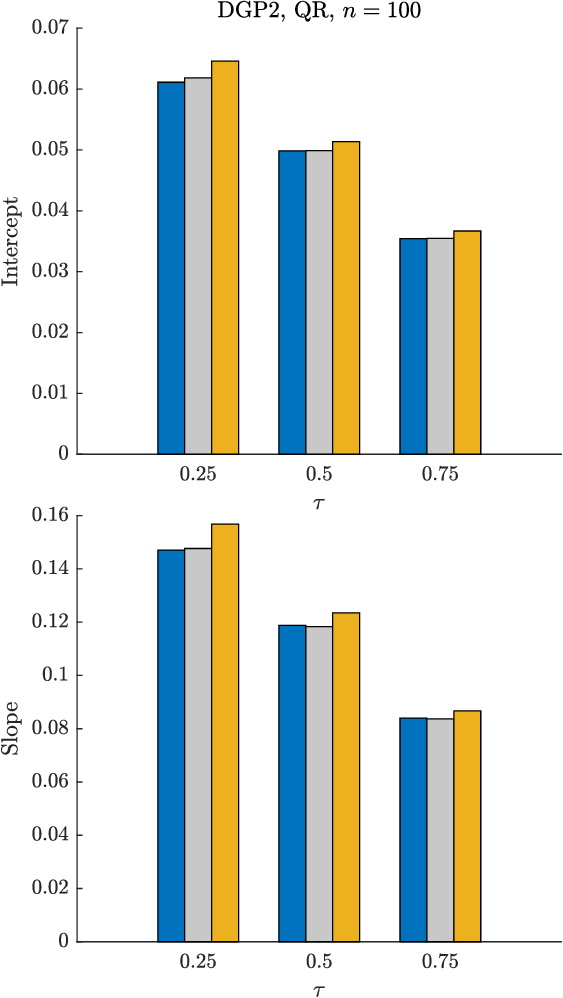}
		\caption{}
	\end{subfigure}
	\begin{subfigure}[b]{0.45\textwidth}
		\includegraphics[width=\textwidth]{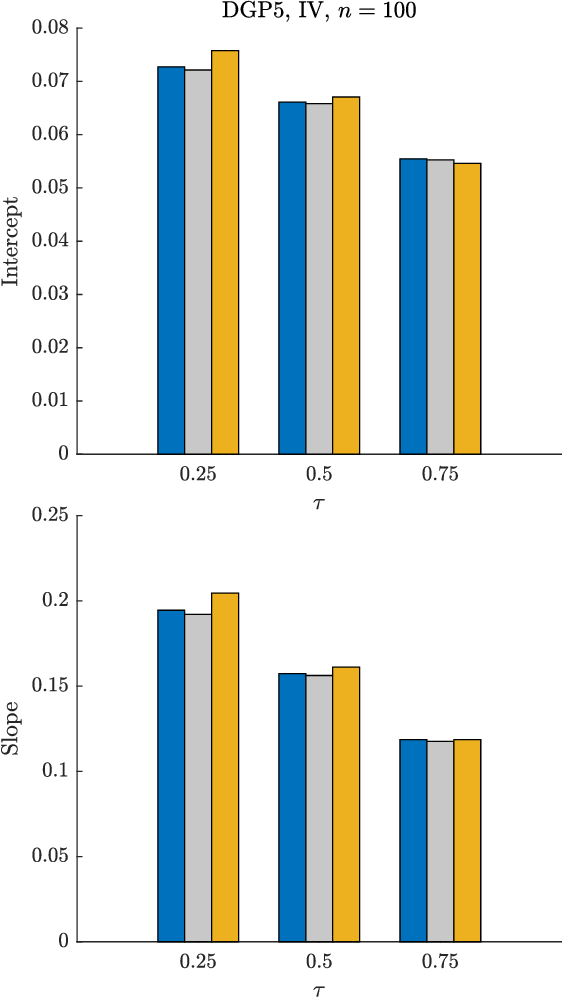}
	\caption{}
	\end{subfigure}

		\end{center}
	 \footnotesize{\textit{Notes:}    The panels compare the RMSE for estimators without bias correction (blue), with infeasible bias correction (grey) and with feasible bias correction based on the rule-of-thumb bandwidth choice (gold) for (a) DGP2, classical QR  and (b) DGP5, IVQR. All results are based on 5,000 simulation repetitions.

	 }
	\label{fig:rmse25}
\end{figure}

\begin{figure}[H]
	\caption{RMSE comparison of raw and bias-corrected estimators }\label{fig:RMSEstable}

	\begin{center}
	\begin{subfigure}[b]{0.45\textwidth}
		\includegraphics[width=\textwidth]{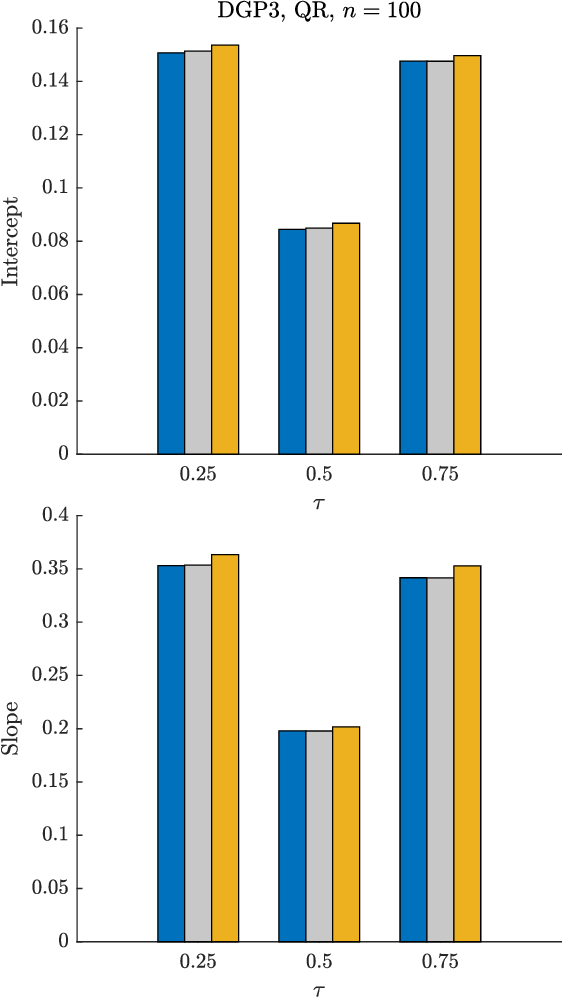}
		\caption{}
	\end{subfigure}
	\begin{subfigure}[b]{0.45\textwidth}
		\includegraphics[width=\textwidth]{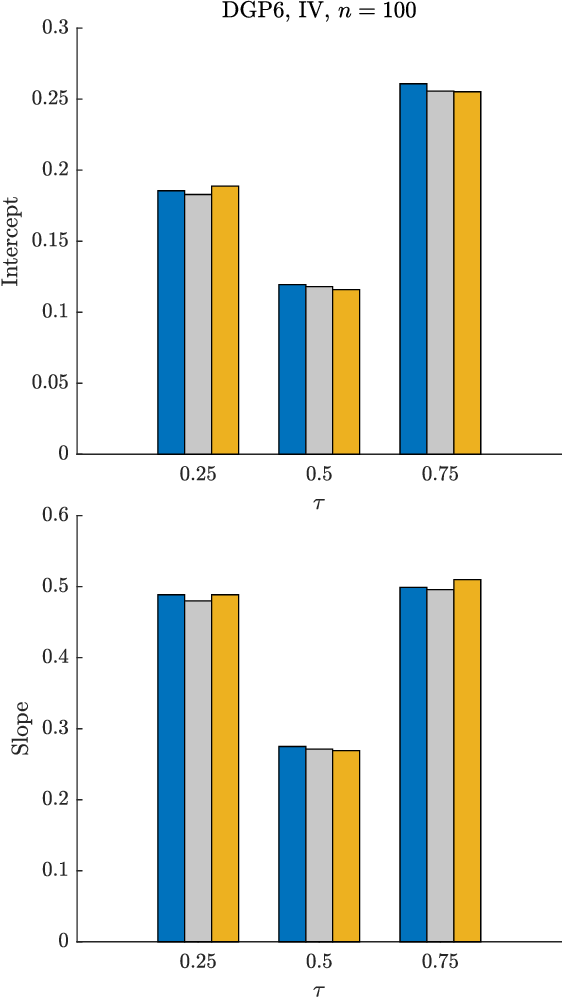}
	\caption{}
	\end{subfigure}

		\end{center}
	 \footnotesize{\textit{Notes:}   The panels compare the RMSE for estimators without bias correction (blue), with infeasible bias correction (grey) and with feasible bias correction based on the rule-of-thumb bandwidth choice (gold) for (a) DGP3, classical QR  and (b) DGP6, IVQR. All results are based on 5,000 simulation repetitions. 		 
	 }
	\label{fig:rmse36}
\end{figure}

\begin{figure}[H]
	\caption{Mean absolute deviations (MAD) of raw and bias-corrected estimators }

	\begin{center}
	\begin{subfigure}[b]{0.45\textwidth}
		\includegraphics[width=\textwidth]{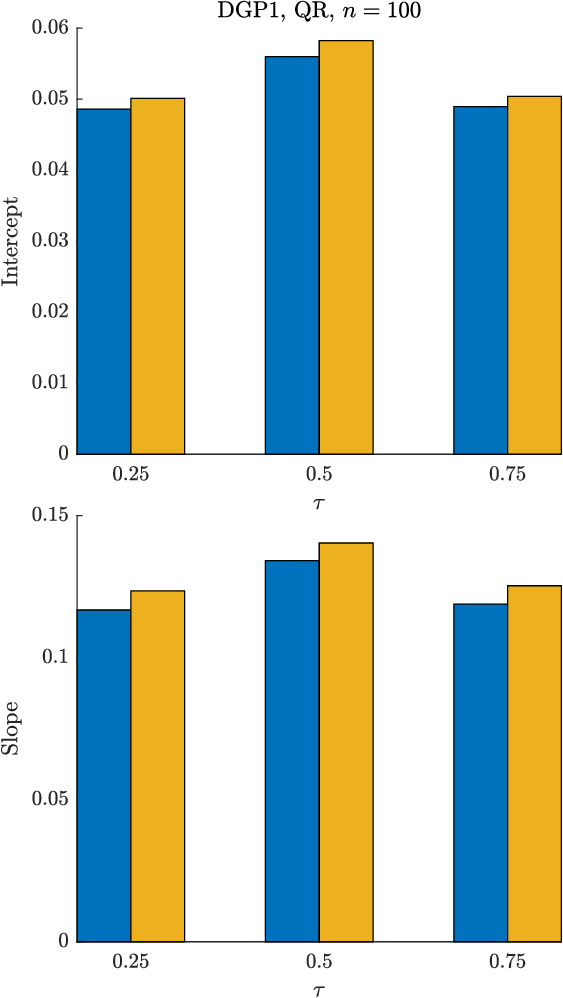}
		\caption{}
	\end{subfigure}
	\begin{subfigure}[b]{0.45\textwidth}
		\includegraphics[width=\textwidth]{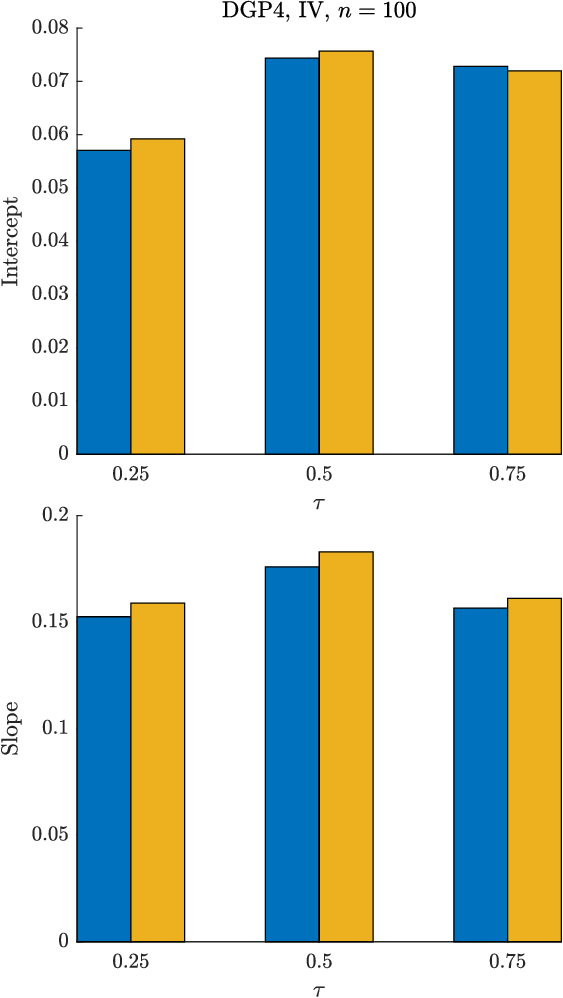}
	\caption{}
	\end{subfigure}

		\end{center}
	 \footnotesize{\textit{Notes:} The panels compare the Mean Absolute Deviation for the estimator without bias correction (blue) and with feasible bias correction (gold) for (a) DGP1, classical QR  and (b) DGP4, IVQR. All results are based on 5,000 simulation repetitions. 	 }
	\label{fig:mad}
\end{figure}

\begin{figure}[H]
	\caption{Confidence interval coverage before and after correction for $n=100$}
	\begin{center}
	\begin{subfigure}[b]{0.3\textwidth}
		\includegraphics[width=\textwidth]{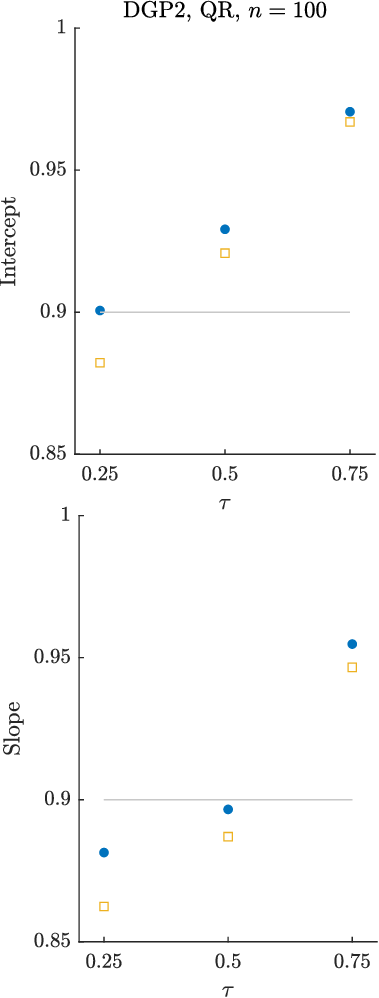}
	\caption{}
	\end{subfigure}
	\begin{subfigure}[b]{0.3\textwidth}
		\includegraphics[width=\textwidth]{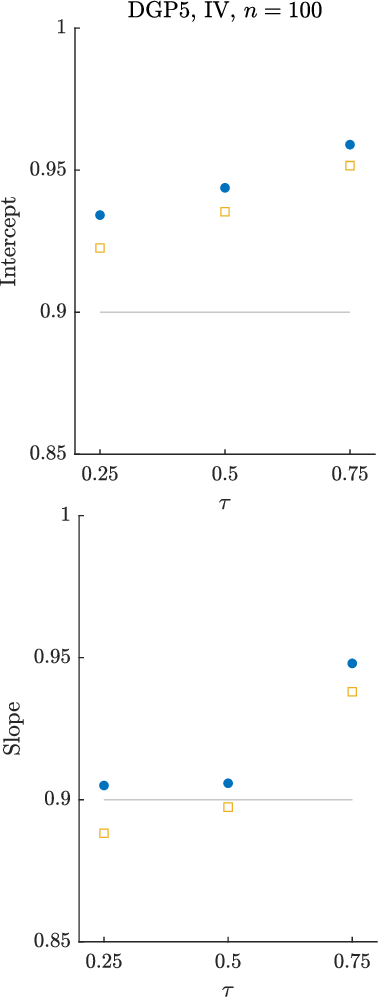}
	\caption{}
	\end{subfigure}

		\end{center}
	 \footnotesize{\textit{Notes:} 
  The panels display the coverage probability of the $90\% $ confidence intervals for the intercept and the slope without bias correction (blue dots) and with the feasible bias correction based on the rule-of-thumb bandwidth choice (gold squares) for DGP2 (classical QR) and DGP5  (IVQR). All results are based on 5,000 simulation repetitions.}
	\label{fig:coverageDG25}
\end{figure}

\begin{figure}[H]
	\caption{Confidence interval coverage before and after correction for $n=100$}
	\begin{center}
	\begin{subfigure}[b]{0.3\textwidth}
		\includegraphics[width=\textwidth]{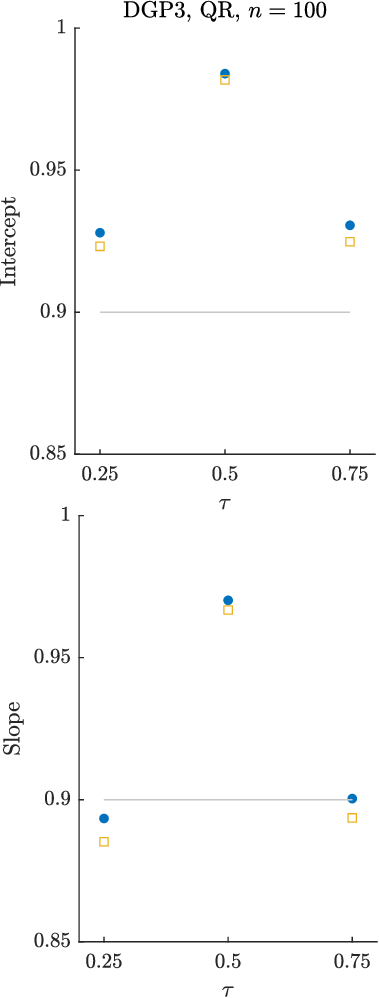}
	\caption{}
	\end{subfigure}
	\begin{subfigure}[b]{0.3\textwidth}
		\includegraphics[width=\textwidth]{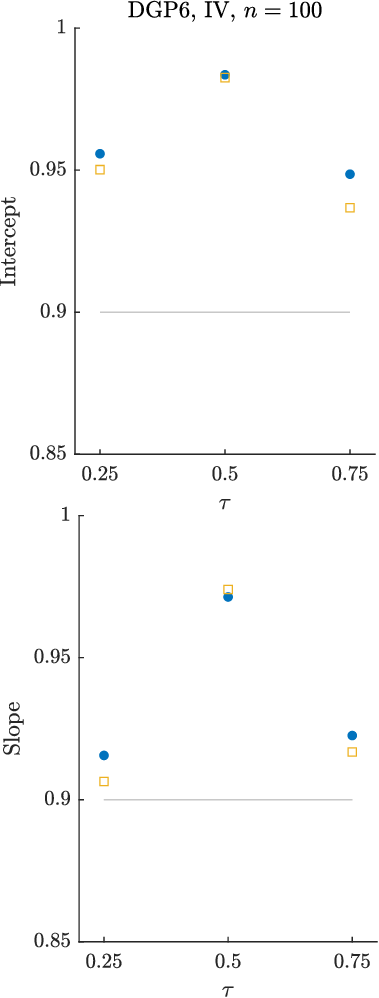}
	\caption{}
	\end{subfigure}

		\end{center}
	 \footnotesize{\textit{Notes:} 
  The panels display the coverage probability of the $90\% $ confidence intervals for the intercept and the slope without bias correction (blue dots) and with the feasible bias correction based on the rule-of-thumb bandwidth choice (gold squares) for DGP3 (classical QR) and DGP6  (IVQR). All results are based on 5,000 simulation repetitions.}
	\label{fig:coverageDG36}
\end{figure}

\begin{figure}[H]
	\caption{Confidence interval coverage before and after correction for $n=200$}
	\begin{center}
	\begin{subfigure}[b]{0.3\textwidth}
		\includegraphics[width=\textwidth]{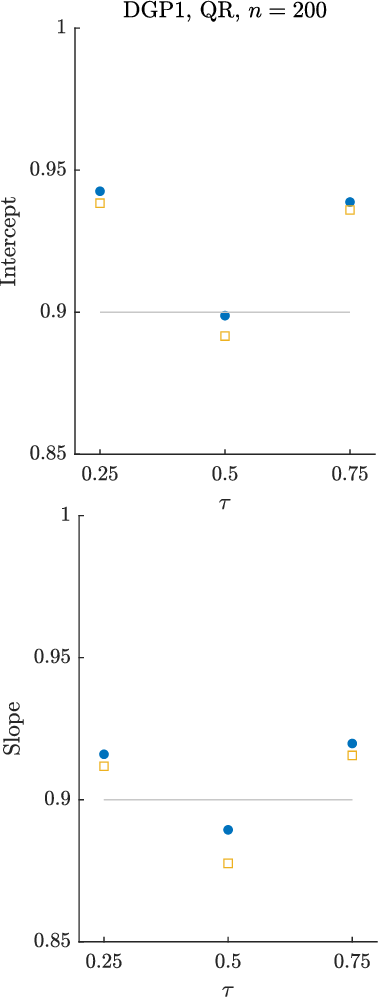}
	\caption{}
	\end{subfigure}
	\begin{subfigure}[b]{0.3\textwidth}
		\includegraphics[width=\textwidth]{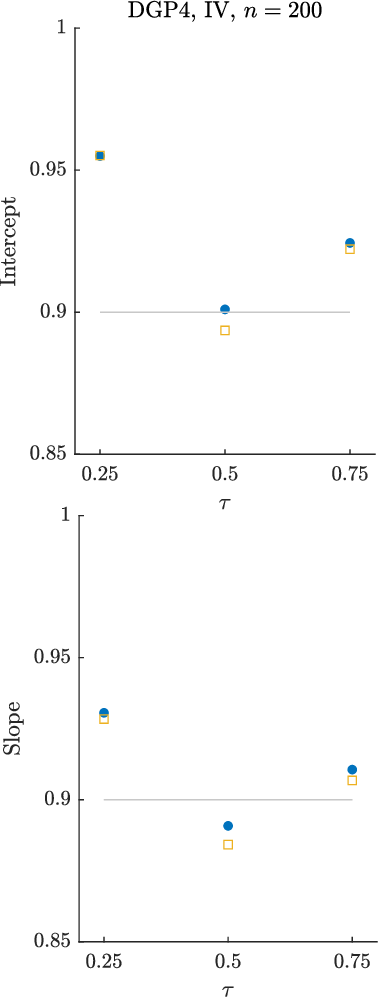}
	\caption{}
	\end{subfigure}

		\end{center}
	 \footnotesize{\textit{Notes:} 
  The panels display the coverage probability of the $90\% $ confidence intervals for the intercept and the slope without bias correction (blue dots) and with the feasible bias correction based on the rule-of-thumb bandwidth choice (gold squares) for DGP1 (classical QR) and DGP4  (IVQR). All results are based on 5,000 simulation repetitions.}
	\label{fig:coverageDGP1_DGP4_n200}
\end{figure}

\end{document}